\documentstyle[preprint,aps,psfig]{revtex}   
\include{amsmath}
\begin{document}

\title{Sturmian theory of three-body recombination: \\
application to the formation of H$_2$ in primordial gas}

\author{Robert C. Forrey}

\address{Department of Physics, Penn State University, 
Berks Campus, Reading, PA 19610-6009}

\date{\today}
\maketitle

\vspace*{-.3in}
\begin{abstract}
A Sturmian theory of three-body recombination is presented 
which provides a unified treatment of bound states, quasi-bound 
states, and continuum states.  The Sturmian representation provides 
a numerical quadrature of the two-body continuum which may be used 
to generate a complete set of states within any desired three-body
recombination pathway. Consequently, the dynamical calculation may 
be conveniently formulated using the simplest energy transfer mechanism, 
even for reactive systems which allow substantial rearrangement.
The Sturmian theory generalizes the quantum kinetic theory of
Snider and Lowry [J. Chem. Phys. {\bf 61}, 2330 (1974)] to include
metastable states which are formed as independent species.
Steady-state rate constants are expressed in terms of a 
pathway-independent part plus a non-equilibrium correction 
which depends on tunneling lifetimes and pressure.
Numerical results are presented for H$_2$ recombination
due to collisions with H and He using quantum mechanical
coupled states and infinite order sudden approximations.
These results may be used to remove some of the uncertainties
that have limited astrophysical simulations of
primordial star formation.

\end{abstract}

\newpage

\section{Introduction}

Three-body recombination (TBR) is one of the most
fundamental types of chemical reaction and has a long
history of study (see \cite{pack1} and references therein).
Probably, the most widely used aproach to computing TBR
rates is the orbiting resonance theory (ORT) developed
by Roberts, Bernstein, and Curtiss (RBC) \cite{rbc}.
In this theory, recombination occurs through a sequential
two-step process where the first step consists of the
formation of a quasibound (QB) orbiting resonance state 
of a two-body subsystem. The second step transfers the
QB state to a bound state through either an energy 
transfer (ET) process or an exchange (Ex) process. For
both two-step mechanisms, the QB states are assumed to
have sufficiently short lifetimes that they maintain   
equilibrium with the free atomic states.

A quantum kinetic theory of chemical recombination was later 
developed by Snider and Lowry \cite{snider} which removed 
the {\it assumption} of equilibrium between the monomers
and dimers. The result appeared to be the same as that of
ORT with the exception that the sum over intermediate QB
states in the ORT formulation be replaced by a complete set 
of intermediate states. Because the intermediate states are 
not considered to be independent molecular species, there
is some flexibility in the choice of pathway for the
TBR process. This flexibility was emphasized by Wei, Alavi, 
and Snider \cite{wei} who formally proved that all pathways 
(including direct 3-body collisions) 
must yield the same rate constant when the sum over two-body states 
is complete, and cautioned that adding partial results from
different pathways could lead to double counting of some of
the transition probability.

In practice, however, the pathway independence has not proven
to be of great utility. Classical dynamical calculations, which
are included in the bulk of TBR studies, generally do not
distinguish between free particle states and the interacting 
continuum. Quantum calculations which do make this distinction
have found it difficult to include broad above barrier (BAB)
resonances and the non-resonant continuum within the ET mechanism 
due to the dynamical requirement that the states be square integrable.
An attempt in this direction was made by Pack, Walker, and 
Kendrick \cite{pack1} who tried to use an $L^2$ representation 
of the BAB resonances. They found unsatisfactory
results because the wave functions for these resonances are less 
localized, and their results depended on the number of nodes that 
were included in the representation. It was concluded that the
BAB resonances and non-resonant continuum could not in practice 
be included in any accurate quantum calculations of the ET mechanism,
and therefore, the contributions from these states to TBR were treated
and handled separately \cite{pack1}.

This conclusion was premature, however, as it was based on fitting, 
truncating, and normalizing $L^2$ functions to numerical wave functions 
obtained from scattering calculations.  If instead the $L^2$ functions
were computed as positive energy eigenfunctions of a Sturmian 
representation of the two-body Hamiltonian, then the apparent 
arbitrariness in truncating the $L^2$ functions found in \cite{pack1}
might be removed. The positive energy eigenfunctions would
provide a quadrature of the continuum which would give a
unified treatment of all QB, BAB, and non-resonant continuum states
and allow convergence tests to be performed.
This approach was used
by Paolini, Ohlinger, and Forrey \cite{paolini} to compute
TBR rates for H$_2$ due to collisions with He and Ar.
The results appeared promising and confirmed the importance
of non-resonant contributions to recombination observed in 
earlier works \cite{pack2,schwenke}.
However, discrepancies remained when the theoretical results 
were compared to existing experimental data. In an effort 
to resolve these discrepancies, it was shown \cite{paolini}
that a model steady-state approximation at high pressure could be 
used to adjust the theoretical results and bring 
them into agreement with the experiment.

In general, a steady-state approximation should be more
accurate than an equilibrium approximation. However, the 
steady-state approximation that is conventionally used \cite{pack1}
and which was used in \cite{paolini} neglects the repopulation of 
intermediate molecules by three-body collisions. In this approximation,
the ORT contribution to recombination rapidly decreases at high
pressure due to a depleted concentration of metastable QB
states.  A master equation analysis performed by Pack, Walker,
and Kendrick \cite{pack2} showed that three-body collisions can
effectively keep the metastable QB states from being depleted
at high pressure. In their study of Ne$_2$ recombination due
to collision with H, they found only a small change ($\sim$7\%) 
in the recombination rate over a large range of pressure.
In the Sturmian theory presented here, we generalize the
theory of Snider and Lowry \cite{snider} to include
metastable states which have lifetimes that are long enough
to survive a three-body collision.
This generalization is formulated in terms of a pathway-independent
contribution plus non-equilibrium corrections which may depend
on the details of the formation. We show that the
pressure dependence completely vanishes for three-body
systems whose internal states are in thermal equilibrium
and confirm the weak dependence on tunneling widths and 
pressure that is found \cite{pack2,schwenke,troe} 
in the more general case. 
The unified treatment of the two-body continuum allows
closed-form expressions to be derived for the non-equilibrium
corrections to the pathway-independent part of the rate constant.
We use the theory to show that the QB contributions to the TBR rate 
do not vanish for large lifetimes as is generally assumed in the
RBC procedure of discarding long-lived QB states \cite{rbc}.
It is also shown that the conventional steady-state approximation 
used in \cite{paolini} to adjust the theory to the experimental 
data is not justified.

A practical benefit of the Sturmian theory is the ability to 
take advantage of the pathway invariance of the TBR rate at equilibrium. 
This allows the dynamical calculation to be formulated in terms 
of the simplest process, which is usually the ET mechanism.
Unlike many calculations in the literature which add the
QB contributions of the ET and Ex mechanisms together
with the direct 3-body contributions, the ET mechanism is 
all that is required in the Sturmian theory.
The Sturmian representation of the intermediate
states provides a numerical implementation of the 
quantum kinetic theory \cite{snider}
which ensures that there are no problems associated with
double counting of the kind described by 
\cite{wei}. 

The theory is applied to the calculation of TBR rates for 
He+H+H and H+H+H, two systems which have been well-studied 
in the past \cite{rbc,whitlock,trainor,jacobs,truhlar,orel,esposito}
but which still have significant uncertainties that limit the 
reliability of current astrophysical simulations of primordial 
star formation \cite{turk}. In order to make use of the
Sturmian representation, we also use a quantum mechanical 
formulation of the three-body dynamics. At the relatively 
high temperatures required by the astrophysical simulations, 
the quantum mechanical formulation is mainly needed for 
identification of the collision complex and for ensuring
that there is no double counting. 
The dimensionality of the quantum mechanical set of
coupled equations may be reduced through angular momentum
decoupling approximations. A comparison of results obtained 
using the coupled states (CS) and infinite order sudden (IOS)
approximations is made in order to determine the temperature
ranges where the angular momentum decoupling is valid.
The results may be used to remove some of the uncertainty 
in the TBR rates that has limited the reliability of the 
astrophysical models.

\section{Theory}

We begin by considering a system of three atoms A, B, and C 
whose internal atomic and molecular states are in thermal 
equilibrium.
We wish to calculate the rate constants
in the effective rate equation
\begin{equation}
[C]^{-1}\frac{d}{dt}[AB]=k_r[A][B]-k_d[AB]
\label{rate}
\end{equation}
where the square brackets denote number density of the enclosed
species, and $k_r$ and $k_d$ are the respective rate constants
for TBR and collision induced dissociation (CID). If the internal
states are not in equilibrium, then the rate constants will not
actually be constant in time, but may be defined as the coefficients
of a steady-state solution to equation (\ref{rate}).
Recombination of two atoms A and B to form molecule AB may
occur through a direct pathway with rate $k_r^0$ or through 
an indirect pathway with rate $k_r^{ET}$ or $k_r^{Ex}$ as shown
\begin{eqnarray}
k_r^0 &:&\hspace{.2in}  A+B+C\rightarrow AB+C
\label{1a}\\
k_r^{ET} &:&\hspace{.2in}  A+B+C\rightarrow A\cdot\cdot\cdot B+C
\rightarrow AB+C
\label{1b}\\
k_r^{Ex} &:&\hspace{.2in}  A+B+C\rightarrow A\cdot\cdot\cdot C+B
\rightarrow AB+C
\label{1c}
\end{eqnarray}
with the corresponding transition operators \cite{wei}
\begin{eqnarray}
T_0 &=& (V-v_{AB})(1+G_E^+\,V)\label{T0}\\
T_{ET} &=& (V-v_{AB})\left[1+G_E^+\,(V-v_{AB})\right]\Omega_{AB}\label{T}\\
T_{Ex} &=& (V-v_{AB})\left[1+G_E^+\,(V-v_{AC})\right]\Omega_{AC}\label{Tx}
\end{eqnarray}
where $G_E^+$ is the outgoing wave Green's function for the full
Hamiltonian consisting of the three-body potential $V$
and the total kinetic energy operator $K_{tot}$. The potentials 
$v_{AB}$ and $v_{AC}$ refer to two-body interactions, and
\begin{equation}
\Omega_{AB}=1+(E^+-K_{tot}-v_{AB})^{-1}v_{AB}
\label{moller}
\end{equation}
is the Moller operator which connects the free continuum with
the interacting continuum.
In the ORT formulation, the $(\cdot\cdot\cdot)$ notation refers to
metastable QB orbiting resonance states, and the recombination rate 
constant is commonly assumed to be $k_r=k_r^0+k_r^{ET}+k_r^{Ex}$.
An additional exchange mechanism may be included for intermediate
$B\cdot\cdot\cdot C$ states if desired. Wei, Alavi, and Snider
have shown \cite{wei} that when the $(\cdot\cdot\cdot)$ notation 
instead refers to a complete set of interacting two-body states, 
the recombination rate constant is $k_r=k_r^0=k_r^{ET}=k_r^{Ex}$.
The Sturmian theory provides such a complete set and allows the
dynamical calculation to be formulated entirely in terms of any
one of the above mechanisms. 

The TBR and CID rate constants may be defined so that the
quantum kinetic theory of Snider and Lowry \cite{snider}
is recovered for a system in local thermodynamic equilibrium (LTE)
\begin{equation}
k_r\equiv\sum_{b,u}k_{u\rightarrow b}\frac{[AB(u)]}{[A][B]}
\hspace{.2in}\Rightarrow\hspace{.2in}
\sum_{b,u}k_{u\rightarrow b}\frac{g_u\exp(-E_u/k_BT)}{Q_AQ_BQ_T}
\ \ \mbox{at LTE}
\label{kr}
\end{equation}
\begin{equation}
k_d\equiv\sum_{b,u}k_{b\rightarrow u}\frac{[AB(b)]}{[AB]}
\hspace{.2in}\Rightarrow\hspace{.2in}
\sum_{b,u}k_{b\rightarrow u}\frac{g_b\exp(-E_b/k_BT)}{Q_{AB}}
\ \ \mbox{at LTE}
\label{kd}
\end{equation}
where $b$ designates a bound state with energy $E_b$ and $u$ designates
an unbound state with energy $E_u$. 
In the Sturmian implementation of equations (\ref{kr}) and (\ref{kd}),
both the bound and unbound states are determined by diagonalizing a two-body
Hamiltonian in an $L^2$ Sturmian basis set with the zero of potential
energy assumed to be at infinite separation. The positive energy eigenstates 
provide a quadrature of the continuum.  If the direct mechanism (\ref{1a})
is used to formulate the dynamics, then $u$ must be a free continuum eigenstate, 
whereas the indirect mechanisms (\ref{1b}) and (\ref{1c}) would require that 
$u$ correspond to an interacting continuum eigenstate.  In order to avoid confusion,
we will generally use $f$ to designate the free continuum states and $u$ to designate 
the interacting continuum states.  The atomic and
molecular partition functions are $Q_A, Q_B,$ and $Q_{AB}$, respectively.
The translational partition function $Q_T$ is defined by
\begin{equation}
Q_T=\frac{1}{h^3}\int_0^{\infty}\exp\left(-\frac{p^2}{2m k_BT}\right)
4\pi p^2dp=\frac{(2\pi m k_BT)^{3/2}}{h^3}
\label{QT}
\end{equation}
where $m$ is the reduced mass of $AB$, $h$ is Planck's constant,
$k_B$ is Boltzman's constant, and $T$ is the temperature.
The discrete sum over unbound states in equations (\ref{kr}) and (\ref{kd})
is a mathematically rigorous approximation for a Sturmian basis set 
whose eigenstates represent quadrature points of the continuum.
This may be verified numerically by using the free energy 
eigenvalues to compute the integral in equation (\ref{QT}) 
\begin{equation}
Q_T=\frac{4\pi m}{h^3}\sum_f \,w_f\,\sqrt{2mE_f}\,\exp(-E_f/k_BT)
\label{QTsturm}
\end{equation}
where $w_f$ are the equivalent quadrature weights \cite{heller}
of the chosen Sturmian representation.
The rate coefficients are defined in the usual way as
\begin{equation}
k_{i\rightarrow j}=\left(\frac{8k_BT}{\pi\mu}\right)^{1/2}
(k_BT)^{-2}\int_0^\infty \sigma_{i\rightarrow j}(E_T)\exp(-E_T/k_BT)\,E_T\,dE_T
\label{kij}
\end{equation}
where $\mu$ is the reduced mass of an atom with respect to a diatom, 
and $E_T=E-E_i$ is the translational energy in the $i$th state, 
which may be taken to be bound or unbound.
For the indirect mechanisms (\ref{1b}) and (\ref{1c}), the collision
cross section $\sigma_{i\rightarrow j}$ refers to non-reactive
and reactive atom-diatom scattering, respectively.
When the transition involves an unbound state, 
the quadrature index enables the cross section, 
which is differential in energy, to exactly handle
energy thresholds and maintain microscopic reversibility.
This allows the principle of detailed balance
\begin{equation}
k_{j\rightarrow i}=\frac{g_i}{g_j}\exp\left(\frac{E_j-E_i}{k_BT}\right)
\,k_{i\rightarrow j}
\label{db}
\end{equation}
to be applied to the LTE limit of equations (\ref{kr}) and (\ref{kd})
which yields the statistical Saha equation 
\begin{equation}
\frac{k_r}{k_d}=\frac{[AB]}{[A][B]}=\frac{Q_{AB}}{Q_AQ_BQ_T}
=(Q_AQ_BQ_T)^{-1}\sum_b g_b\exp(-E_b/k_BT)
\label{saha}
\end{equation}
for the thermalization of the continuum.

For a system which is not in thermal equilibrium, a more detailed
rate analysis is required. The effective rate equation (\ref{rate})
may be replaced by a set of state-specific rate equations
\begin{eqnarray}
\frac{d}{dt}[AB(b)] &=& [C]\sum_u
\left(k_{u\rightarrow b}[AB(u)]-k_{b\rightarrow u}[AB(b)]\right)\nonumber\\
&+& [C]\sum_{b'}\left(k_{b'\rightarrow b}[AB(b')]-k_{b\rightarrow b'}[AB(b)]\right)
\label{rateb}
\end{eqnarray}
\begin{eqnarray}
\frac{d}{dt}[AB(u)] &=& [C]\sum_b
\left(k_{b\rightarrow u}[AB(b)]-k_{u\rightarrow b}[AB(u)]\right)
+k_{f\rightarrow u}^{elastic}[A][B]\nonumber\\
&+& [C]\sum_{u'} \left(k_{u'\rightarrow u}[AB(u')]-k_{u\rightarrow u'}[AB(u)]\right)
-\tau^{-1}_{u}[AB(u)]
\label{rateu}
\end{eqnarray}
where $\tau_{u}$ is the lifetime of the unbound state $u$,
and $k_{f\rightarrow u}^{elastic}$ is the two-body elastic scattering
rate constant for the reverse process which may be computed from $\tau_{u}^{-1}$ 
using the same equilibrium constant 
\begin{equation}
K_u^{eq}=\frac{g_u\exp(-E_u/k_BT)}{Q_AQ_BQ_T}
\end{equation}
introduced in the LTE limit of equation (\ref{kr}).
The steady-state solution to equation (\ref{rateu}) is given by
\begin{equation}
\frac{[AB(u)]}{[A][B]}=\frac{K_u^{eq}
+\tau_u[C][A]^{-1}[B]^{-1}\left(\sum_b k_{b\rightarrow u}[AB(b)]
+\sum_{u'\neq u} k_{u'\rightarrow u}[AB(u')]\right)}
{1+\tau_u[C]\left(\sum_b k_{u\rightarrow b}
+\sum_{u'\neq u} k_{u\rightarrow u'}\right)}
\label{stateu}
\end{equation}
In the conventional steady-state approximation,
the CID contribution $k_{b\rightarrow u}$ is assumed to be
small due to the excitation threshold, and the $k_{u'\rightarrow u}$
contribution in the numerator of equation (\ref{stateu}) is
neglected for no particularly good reason. It is precisely
this term that Pack, Walker, and Kendrick \cite{pack2}
discovered in their master equation analysis which
keeps the QB states from being depleted at high pressures.
When it is neglected, the numerator of equation (\ref{stateu})
would approximately equal $K_u^{eq}$ and $[AB(u)]$ would appear
to be very small for long-lived QB states and high concentrations $[C]$.
When this steady-state result is used to obtain the ORT 
recombination rate, there would be a substantial falloff at high pressures. 
As noted in \cite{pack2}, such strong ORT ``falloff" is not generally
observed in experiments.

A better approach is to solve the rate equations
(\ref{rateb}) and (\ref{rateu}) to obtain steady-state concentrations.  
The quantum kinetic theory \cite{snider} already accounts for steady-state
behavior arising from the transient formation and decay of interacting
continuum states. The result is an equilibrium concentration of all 
bound and unbound states. A complete set of unbound states includes 
both the resonant contributions of ORT and the non-resonant contributions 
which help to maintain the equilibrium concentrations.
Perturbations from equilibrium may occur for long-lived QB states.
In order to generalize the theory to 
include non-LTE behavior, it is convenient to 
write equations (\ref{kr}) and (\ref{kd}) as
\begin{equation} 
k_r\equiv\sum_{b,u}(1+\delta_u)k_{u\rightarrow b}
\frac{g_u\exp(-E_u/k_BT)}{Q_AQ_BQ_T}
\label{krnew}
\end{equation}
\begin{equation}
k_d\equiv\sum_{b,u}(1+\delta_b)k_{b\rightarrow u}
\frac{g_b\exp(-E_b/k_BT)}{Q_{AB}}
\label{kdnew}
\end{equation}
and express the solution of the rate equations
in terms of the pathway-independent part plus a
term containing the non-LTE concentration defects 
\begin{equation}
\delta_u=\frac{\tau_u[C]\left(\sum_b \delta_b\, k_{u\rightarrow b}
+\sum_{u'\neq u} \delta_{u'}\, k_{u\rightarrow u'}\right)}
{1+\tau_u[C]\left(\sum_b k_{u\rightarrow b}
+\sum_{u'\neq u} k_{u\rightarrow u'}\right)}
\label{deltau}
\end{equation}
\begin{equation}
\delta_b=\frac{\sum_{b'} \delta_{b'}\,\Gamma_{b\rightarrow b'}
+[C]\left(\sum_u \delta_u\, k_{b\rightarrow u}
+\sum_{b'\neq b} \delta_{b'}\, k_{b\rightarrow b'}\right)}
{\sum_{b'} \Gamma_{b\rightarrow b'}
+[C]\left(\sum_u k_{b\rightarrow u}+\sum_{b'\neq b} k_{b\rightarrow b'}\right)}
\label{deltab}
\end{equation}
where $\Gamma_{b\rightarrow b'}$ has been added to allow for radiative transitions.
Note that the LTE part of the rate constants (\ref{krnew}) and (\ref{kdnew})
does not depend on pressure. The non-LTE concentration defects allow for QB states
that have lifetimes that are long enough to survive a three-body collision 
or are formed through an independent process. 
The scale invariance of the homogeneous equations (\ref{deltau}) and (\ref{deltab})
is due to microscopic reversibility. The choice of scale may be determined by
the physical conditions under consideration. The defects are negative for states 
which are underpopulated relative to a thermal distribution and are small in 
magnitude unless there is an efficient relaxation pathway which competes with 
the collisions. 
%These defects are negative for states which are underpopulated relative 
%to a thermal distribution and are small in magnitude unless there is an 
%efficient relaxation pathway which competes with the collisions. 
Equations (\ref{deltau}) and (\ref{deltab}) refer to the ET mechanism, 
but it is here that any important exchange mechanisms may also be included 
in the kinetics (see below). 

Equation (\ref{deltau}) has much in common with the kinetic model
proposed in \cite{pack2}.  At low pressures and short lifetimes, 
the $\delta_u$ makes a neglible contribution and the recombination 
rate is simply equal to the pathway-independent LTE result. Note 
that in this limit, the QB states still contribute when an indirect
mechanism is used to compute the rate constant. 
At high pressure,  the $\delta_u$ becomes important for QB states with
large lifetimes.  Equation (\ref{deltau}) shows that when the defects
$\delta_b$ and $\delta_{u'}$ are negative, the pressure-dependent QB 
contribution gets subtracted from the pathway-independent part. 
This provides a small ``falloff" to the recombination rate of 
the kind observed in \cite{pack2} where the low and high pressure
limits differed by only $\sim7\%$. This small falloff compared to ORT
is due to a combination of the inclusion of short-lived $u$-states and 
the incomplete removal of long-lived $u$-states in equation (\ref{krnew}).

A similar result is obtained when the system allows
a QB state of the $A\cdot\cdot\cdot C$ complex.
In this case, the $\delta_u$ in equations (\ref{krnew}) and
(\ref{deltau}) are simply replaced by 
\begin{equation}
\delta_u^{AB}=\frac{\tau_u^{AB}[C]\left\{
\sum_b (\delta_b^{AB}\, k_{u\rightarrow b}^{ET}
+\delta_b^{AC}\, k_{u\rightarrow b}^{Ex})
+\sum_{u'\neq u} (\delta_{u'}^{AB}\, k_{u\rightarrow u'}^{ET}
+\delta_{u'}^{AC}\, k_{u\rightarrow u'}^{Ex})\right\}}
{1+\tau_u^{AB}[C]\left\{\sum_b (k_{u\rightarrow b}^{ET}
+k_{u\rightarrow b}^{Ex})+\sum_{u'\neq u} (k_{u\rightarrow u'}^{ET}
+k_{u\rightarrow u'}^{Ex})\right\}}
\label{delta_ab}
\end{equation}
with the superscripts AB and AC used to clarify the lifetime
and concentration defects, and the superscripts ET and Ex
used to denote a direct or exchange collision. This expression
is easily generalized for systems which also allow a QB state 
of the $B\cdot\cdot\cdot C$ complex. The key point is that
the non-LTE corrections are where the dependence on formation 
pathway, tunneling lifetimes, and density should be included 
in the calculation of the rate constants $k_r$ and $k_d$.
Any pressure dependence which might arise from pathways
that are different from the one used to compute the
LTE part should not be incoherently added together
as is commonly done in the literature. 
Instead, such contributions should be subtracted from the 
LTE part with weights determined by the non-LTE defects.
Equations (\ref{deltau}), (\ref{deltab}),
and (\ref{delta_ab}) provide closed-form expressions
which may be solved self-consistently, thereby removing
the need for any further kinetic considerations. This
approach is far simpler and more transparent than the 
alternative method \cite{pack2,schwenke} of numerically 
solving the master equations 
and then fitting the results to the effective rate equation (\ref{rate}).
It is also less prone to errors that may occur when using cross sections 
which do not exactly satisfy detailed balance \cite{pack2}.
Because the pathway-independent part of the rate constant
already accounts for much of the kinetics, the non-LTE defects
will be small unless there is an efficient mechanism which allows 
external energy input. Equations (\ref{deltau}) and (\ref{deltab}) 
may be used to define critical densities
\begin{equation}
[C]_{cr}^{\tiny{\mbox{falloff}}}=\frac{1}{\tau_u\left(\sum_b k_{u\rightarrow b}
+\sum_{u'\neq u} k_{u\rightarrow u'}\right)}
\label{cr1}
\end{equation}
\begin{equation}
[C]_{cr}^{\tiny{\mbox{LTE}}}=\frac{\sum_{b'} \Gamma_{b\rightarrow b'}}
{\sum_u k_{b\rightarrow u}+\sum_{b'\neq b} k_{b\rightarrow b'}}
\label{cr2}
\end{equation}
where pressure falloff and departures from LTE would be expected
to occur. These critical densities characterize competing processes.
For example, when $[C]>[C]_{cr}^{\tiny{\mbox{falloff}}}$ for a particular
QB state, one might expect $\delta_u\approx -1$ so that the contribution 
from this QB state would be significantly reduced. This would be consistent
with the RBC procedure of removing such long-lived QB states from the ORT
recombination rate. However, if the condition $[C]>[C]_{cr}^{\tiny{\mbox{LTE}}}$
is also satisfied for a particular bound state, then $\delta_b\approx 0$ for
this bound state. This would tend to move the $\delta_u$ closer to zero
and reduce the pressure falloff. 
%This issue is studied further in section IV 
Section IV gives estimates of critical densities and shows
that the pressure falloff is generally small for an isolated H$_2$ system.
These estimates are based on rate coefficients obtained from quantum mechanical 
calculation of the collision cross sections as described below.

A quantum mechanical calculation of the collision
cross section needed in equation (\ref{kij}) may be
obtained by considering the atom-diatom Hamiltonian
in the center of mass frame
\begin{equation}
H=-\frac{1}{2m}\,\nabla^2_r-\frac{1}{2\mu}\,\nabla^2_R
+v_{AB}(r)+V_I(r,R,\theta)
\label{ham}
\end{equation}
where $r$ is the distance between atoms $A$ and $B$,
$R$ is the distance between atom $C$ and the center of mass
of $A\cdot\cdot\cdot B$, $\theta$ is the angle between $r$ and $R$,
$m$ and $\mu$ are defined as above, and the three dimensional
potential energy surface (PES) in equations (\ref{T0})-(\ref{Tx})
is separated into a diatomic potential $v_{AB}(r)$ and an interaction
potential $V_I(r,R,\theta)$ for the ET mechanism.
Perhaps the most significant advantage in using the 
Sturmian theory to compute TBR rates is that it
allows the pathway-independent part of the rate coefficients 
(\ref{krnew}) and (\ref{kdnew}) to be calculated entirely
in terms of the ET mechanism. The dynamics for these
rate coefficients then 
reduces to non-reactive scattering calculations
which are generally easier to deal with than
the reactive scattering calculations that would
be needed to formulate the theory in terms
of the Ex mechanism.
In cases where there is strong interference
between reactive and non-reactive channels,
this may not amount to much of an advantage
as any quantum dynamical calculation would
require a full account of all arrangement channels. 
This would likely be the case for attractive systems
at low temperatures. However, for temperatures that are
high enough that classical trajectories would
permit a good approximation to the dynamics, 
any quantum interference between the reactive
and non-reactive channels would be unimportant,
and the non-reactive channels could be separated out. 
In this case, it is still desirable to perform 
quantum dynamical calculations
in order to utilize the Sturmian basis set
and avoid the double counting problems
discussed by Wei, Alavi, and Snider \cite{wei}.

Before turning to a numerical application of the theory,
it is useful to summarize the various sources of uncertainties
that may arise in the computation of the TBR and CID rate constants.
These uncertainties include:
(i) an accurate accounting of the kinetic pathways,
(ii) an accurate accounting of thermodynamic variables,
(iii) numerical convergence of the dynamical solutions 
with respect to basis set size,
(iv) level of decoupling approximation, and
(v) accuracy of the potential energy functions.
The Sturmian theory allows complete control of 
(i)-(iii) as described above. For identical 
particles, the ET and Ex mechanisms can be
included together with the proper symmetrization,
which would reduce the basis set size and the
corresponding uncertainty (iii). Uncertainties
(iv) and (v) are the primary sources of uncertainty
for the application involving the formation of H$_2$ 
considered below.

\section{Application}

Astrophysical simulations of primordial star formation 
require TBR rate constants for H$_2$ formation as input.
Published rate constants show orders of magnitude disagreement
at temperatures required by the simulations, and it was
concluded in a recent study \cite{turk} that the uncertainty 
in the TBR rate ``represents a major limitation on our ability
to accurately simulate the formation of the first stars in the 
universe."
In a recent paper \cite{bob}, we reported that a factor of 
$\sim 100$ uncertainty which was introduced by three of the most
commonly-used rate constants \cite{turk} may be reduced 
to a factor of $\sim 2$ by the Sturmian theory. 
Subsequent simulations \cite{bovino} have used
our calculated rate constant to study the
gravitational collapse of primordial gas.
While the factor of 2 uncertainty that we reported is not 
rigorously proven, we show here the details of the calculation
and the reasoning behind the estimate. As alluded to above, 
the calculations are formulated using the ET mechanism.

The first step of the ET mechanism converts the free Sturmian
eigenstates to interacting Sturmian eigenstates as prescribed
by the Moller operator (\ref{moller}).
This requires solution of the diatomic Schr\"odinger equation 
\begin{eqnarray}
\left[\frac{1}{2m}\,\frac{d^2}{dr^2}-\frac{j(j+1)}{2m\,r^2}
-v_{AB}(r)+E_{vj}\right]\,\chi_{vj}(r)=0\ ,
\label{diatom}
\end{eqnarray}
where $v$ and $j$ are the vibrational and rotational quantum
numbers for the eigenstate $\chi_{vj}$.
The bound ro-vibrational wave functions and discretized positive energy
states are obtained by diagonalization of the diatomic Hamiltonian in an
orthonormal $L^2$ Sturmian basis set. 
For light atomic systems, such as the hydrogen pairs considered here,
it is convenient to use a Sturmian representation consisting of
Laguerre polynomial $L_n^{(2j+2)}$ basis functions
\begin{eqnarray}
%\phi_{l,n}(r)&=& \sqrt{\frac{a n!}{(n+2l+2)!}}
%\,(ar)^{l+1}\exp(-ar/2)L_n^{(2l+2)}(ar)\ .
\phi_{j,n}(r)&=& \sqrt{\frac{a n!}{(n+2j+2)!}}
\,(ar)^{j+1}\exp(-ar/2)L_n^{(2j+2)}(ar)
\end{eqnarray}
where the scale factor $a$ plays an important role in
controlling the convergence rate as shown in section IV.
The notation $b$ and $u$ in section II is now understood to mean
a bound or unbound eigenstate which is characterized by the pair
of quantum numbers $(v,j)$. Note that the vibrational quantum number 
for an unbound eigenstate corresponds to the quadrature index and only 
has meaning with respect to the scale factor and number of basis functions.

The second step uses the transition operator $T_{ET}$ 
in equation (\ref{T}) to compute the cross section.
Many techniques have been developed for this purpose, 
so here we provide only a brief overview.
In the close-coupling (CC) method \cite{lester},
the full wave function is expanded in terms of 
channel functions characterized by an index $m\equiv\{v,j,l\}$ 
which leads to a set of coupled equations of the form
\begin{eqnarray}
\left[\frac{d^2}{dR^2}-\frac{l_{m}(l_{m}+1)}{R^2}
+k_{m}^2\right]\,C_{m}(R)
=2\mu\sum_{n}\,C_n(R)\,\langle m|V_I|n\rangle
\label{cc}
\end{eqnarray}
where $l_m$ is the orbital angular momentum in the $m$-th channel
and $k_{m}^2=2\mu(E-E_m)$ is the square of the translational wave number.
Because the positive energy eigenstates in the Sturmian representation
correspond to quadrature points of the two-body continuum \cite{heller},
the uncountably infinite set of continuum states is coupled together
in an approximation scheme which mirrors what 
is usually done for calculating transitions
%between bound states. Therefore, it should be possible to compute TBR
between bound states. Therefore, in principle it is possible to compute TBR
rate coefficients using the numerically exact CC method for systems
which do not require coupling to states with large values of $j$.
The dimensionality of the set of coupled equations grows
rapidly with increasing $j$ due to the exact treatment of the
angular momentum coupling, so it is also desirable to consider
various decoupling approximations. One of the most reliable 
decoupling approximations is the coupled states or centrifugal
sudden (CS) approximation \cite{pack,kouri,mcguire} 
which replaces the $m$-th channel orbital angular momentum
$l_m$ with an average value $\overline{l}$ and reduces the
channel index to include only the state quantum numbers $v$ and $j$.
The infinite-order-sudden (IOS) approximation \cite{parker} makes the 
additional approximation that the internal rotational energy is averaged 
over so that the orientation angle $\theta$ is treated as a parameter.
A more severe IOS approximation \cite{pfeffer} averages over both the 
internal rotational and vibrational energy so that the set of coupled
equations (\ref{cc}) reduces to a set of uncoupled one-dimensional 
equations. This version was used by Pack et al. \cite{pack1} to study
TBR of Ne$_2$ due to collisions with H, and it was estimated to be accurate to 
within about 20\% at low temperatures ($\sim 30$ K) due to the closely spaced 
energies.  The internal energies for H$_2$ are not as closely spaced, however,
the spacing decreases with excitation, and for the most important 
transitions to highly excited states, the approximation is
expected to give semiquantitative accuracy \cite{sakimoto} which improves 
as the translational temperature increases. The IOS results
reported in this work correspond to this most simplified version of
the approximation. 

Various reactive IOS approximations have also been developed \cite{khare}
and applied to H+H$_2$ collisions \cite{kouri2,bowman,opdehaar,light}.
In the reactive IOS with optical potential \cite{baer},
the problem reduces to a non-reactive formulation
for a single arrangement. Similarly, an $L^2$ Sturmian representation 
of the ET mechanism is able to convert a multi-arrangement system into 
an inelastic single arrangement system. However, negative imaginary potentials
cannot be used to absorb large-$r$ flux due to the free three-body boundary 
condition. Instead, the convergence of the Sturmian method relies on the 
square integrability of the basis set. For reactive systems, the CC and
CS formulations typically require that all basis functions used to
represent the free and interacting continuum be coupled together
in the dynamics.  This will in many cases present an insurmountable
practical limitation. The IOS approximation, however, decouples
the internal coordinates and allows a practical solution
for cases where multiple scattering and interference effects
are expected to be small.
%Furthermore, when the Sturmian basis set is used within the IOS
When the Sturmian basis set is used within the IOS
approximation, the energy thresholds for the unbound states are 
exactly handled and the cross sections exactly satisfy the principle 
of microscopic reversibility. Therefore, the rate coefficients
exactly satisfy detailed balance, which is essential to
the kinetic analysis given above. This differs from the
implementation of the IOS approximation used in \cite{pack1}
where the rate coefficients did not completely satisfy
detailed balance.

The principle of detailed balance has been a key ingredient
in determining TBR and CID rate constants for the formation
and destruction of H$_2$ in astrophysical environments,
and there has been some confusion in the literature about 
how it should be applied to phenomenological rate constants 
such as $k_r$ and $k_d$ \cite{bob}.
Three of the most commonly used TBR rates vary by more than 100 
times at temperatures needed to simulate primordial star formation \cite{turk}. 
Two of these rates are derived from the same experimental data of 
Jacobs et al. \cite{jacobs} and their differences are due to
differences in the application of detailed balance \cite{bob}.
All of the rate constants used in the simulations \cite{turk} 
rely on extrapolations for the temperature range 300-2900 K.
The Sturmian theory is used in the next section to compute
the rate constant for H$_2$ formation in this temperature range.
The quantum mechanical calculations are compared to measurements 
of Trainor et al. \cite{trainor} for He colliders
and Jacobs et al. \cite{jacobs} for H colliders.
We also compare our results with the DEB quasiclassical 
calculations of Esposito and Capitelli \cite{esposito}.
Their DEB label refers to detailed balance applied to 
direct CID from bound states. Classical calculations 
do not distinguish between the free and interacting continuum, 
and it was assumed that direct dissociation involves exclusively
the free continuum \cite{esposito}. If this interpretation is correct,
then upon application of detailed balance, the DEB curve would 
correspond most closely to the rate constant $k_r^0$, which
we have argued is the same as the pathway-independent part 
of the effective rate constant $k_r$. 

In separate calculations, Esposito and Capitelli 
performed classical dynamics calculations of the ORT
contributions for both the ET and Ex mechanisms \cite{esposito}.
The pressure variation of these calculations was shown to be 
about an order of magnitude over the temperature range 300-3000 K. 
This variation was due to a kinetic scheme, similar to 
the RBC procedure, which selects QB states based on their associated 
lifetimes.  We show below that this kinetic scheme, like the conventional
steady-state approximation used in \cite{paolini}, is not
justified for H$_2$ formation. Three-body collisions are
%able to maintain an equilibrium population of QB states for this system,
%so the TBR rate does not vary with pressure under normal laboratory conditions.
able to maintain an equilibrium population of QB states for this system
under normal laboratory conditions. Therefore, the pressure dependence
should be much less than what was found from the ORT contributions
using their kinetic model \cite{esposito}.

The DEB and ORT results were added together assuming a 
temperature over pressure ratio of 3000 (which corresponds 
to [H]=$2.4\times10^{18}$ cm$^{-3}$) and improved agreement 
with the experiment of Jacobs et al \cite{jacobs} 
was found \cite{esposito}. The improvement was slight
due to the smallness of the ORT contribution in the
experimental temperature range 2900-4700 K at the assumed
pressure.  The ORT result \cite{esposito}, however, 
increases with decreasing pressure and temperature, and 
would be significant for the conditions required in the 
astrophysical simulations. Based on our analysis, these two 
contributions should not be added together. At the low
densities of the primordial gas, there may be a small
pressure dependence due to a non-LTE population of
excited bound and QB states. However, in this case
the ORT result with an appropriate weighting would 
need to be {\em subtracted} from the pathway-independent 
LTE rate constant in order to remove the long-lived QB states 
which are under-populated.

\section{Results}

The pathway-independent part of the TBR rate constant (\ref{krnew})
is calculated for He+H+H and H+H+H using the ET mechanism. In both
cases, the H$_2$ potential is obtained using a fit \cite{schwenke1}
to the Ad potential of Schwenke \cite{schwenke2}. The MR PES \cite{mr}
and also an additive pair potential PES were separately used for the 
HeH$_2$ system. The BKMP2 PES \cite{bkmp2} was used for the H$_3$ system.
%In this work, we included $n=\{0,...,100\}$ for each value
For the Sturmian basis set used in this work, we included $n=\{0,...,100\}$ 
for each value of $j\leq 35$. The diatomic Schr\"odinger equation (\ref{diatom})
was solved numerically using this basis set for both the free
particle interaction $v_{AB}=0$ and the full two-body interaction for H$_2$.
The accuracy of the Sturmian representation may be assessed in part by
its ability to reproduce the translational partition function $Q_T$
using the free particle energy eigenvalues as given by equation (\ref{QTsturm}). 
The equivalent quadrature weights $w_f$ were computed as numerical
derivatives of the discrete energy eigenvalues with respect to 
index number \cite{heller} using a spline fit.
Figure 1a shows the percent error between equations 
(\ref{QT}) and (\ref{QTsturm}) for different values
of scale parameter $a$. The Sturmian eigenvalues
in this plot are associated with $j=4$, however,
the pattern is similar for all $j$.
For the 101 basis functions used here, 
the accuracy of the Sturmian representation of $Q_T$
is seen to be best for the smallest scale factor shown. 
The accuray of the interacting eigenvalues, whose bound 
and QB states are governed by a variational principle, 
is also affected by the choice of scale parameter, so 
an appropriate balance between the two is desirable.
In the present work, all of the bound and QB energies
obtained from the diagonalization agree with those
reported by Schwenke \cite{schwenke2} to within 5\%. 
The tunneling widths and lifetimes indicated below
are also taken from \cite{schwenke2}.
Figure 1b shows the
percent error using equation (\ref{QTsturm}) with the 
positive energy eigenvalues of the interacting continuum.
The calculations were performed both with and without the
%long-lived $j=4$ QB state ($\Gamma\sim 10^{-5}$ cm$^{-1}$)
long-lived $j=4$ QB state ($\tau\sim 6\times 10^{-7}$ sec)
included in the summation. At high temperatures, the 
interacting continuum which includes the QB state in
the summation gives a better approximation to the
translational partition function.

Figure 2 shows the 14th vibrational eigenstate for a
$j=4$ Sturmian representation of the free and interacting 
continuum of H$\cdot\cdot\cdot$H using a scale parameter $a=20$.
%Note that the vibrational quantum number for a positive energy eigenstate
%corresponds to the quadrature index and only has meaning with respect to 
%the scale parameter.
In both cases, the eigenstate is plotted
versus interatomic distance using the normalization
\begin{equation}
\psi=\frac{\chi_{vj}(r)}{\sqrt{w_{vj}}}
\hspace{.2in}\Rightarrow\hspace{.2in}
\sqrt{\frac{2m}{\pi k_{vj}}}\,k_{vj}r\,j_j(k_{vj}r)
\ \ \mbox{for free continuum}
\label{norm}
\end{equation}
where $k_{vj}=\sqrt{2mE_{vj}}$ and $j_l$ is a spherical Bessel function.
As may be seen in Figure 2a, the Sturmian representation shows excellent
agreement with the exact free continuum state for $r\le 20$ a.u. before
going quickly to zero as required by the exponential term in the basis functions.
All of the free and interacting continuum states are cut off at 
the same distance. This cut-off distance is not an arbitrary choice, 
but is instead controlled by the choice of scale parameter. 
The interacting eigenstate shown in Figure 2b corresponds to a long-lived 
QB state with tunneling width $\Gamma\sim 8\times 10^{-6}$ cm$^{-1}$
and resonant energy $E_r\sim 0.9$ cm$^{-1}$.

Another long-lived QB state is shown in Figure 3 for $j=15$
and $v=13$. Also shown in the figure are wavefunctions for
the non-resonant $v=12$ and $v=14$ states.  The QB state 
($\Gamma\sim 3\times 10^{-6}$ cm$^{-1}$ and $E_r\sim 190$ cm$^{-1}$)
is localized below 10 a.u. which produces a strong contribution to 
the TBR rate.  Clearly, there is neglible overlap between the QB state 
and the neighboring continuum.  Higher vibrational levels do show more 
significant overlap and can help keep the QB concentration from being 
depleted at high pressures.  The pressure dependence of this resonant 
contribution to the TBR rate is examined below.

Figure 4 shows the cumulative bound-free transition probability
as a function of the positive energy Sturmian eigenvalues.
The calculations are for He+H$\cdot\cdot\cdot$H$(j=4)$ 
computed in the IOS approximation using the MR PES \cite{mr}. 
The curves shown in the figure include a summation of the
state-to-state probabilities over all bound states, and 
each point was computed with $E_T^{(u)}=100$ cm$^{-1}$
and $\overline{l}_{max}=30$.  The state-to-state probabilities, 
which are defined as
\begin{equation}
P_{b\leftrightarrow u}(E)
=\left(\frac{2\mu E_T^{(b)}}{\pi}\right)
g_b\,\sigma_{b\rightarrow u}(E_T^{(b)})
=\left(\frac{2\mu E_T^{(u)}}{\pi}\right)
g_u\,\sigma_{u\rightarrow b}(E_T^{(u)})
\end{equation}
have a scale dependence when computed using the Sturmian
eigenstates. This is due to the unit normalization of the
$L^2$ basis functions. 
The Sturmian eigenstates could easily be energy-normalized
as was done in equation (\ref{norm}).
However, we are only interested in the sum of the probabilities 
over $u$-states, so it is not necessary to energy-normalize these 
states. 
The weights $w_f$ needed for a quadrature of the free continuum
cancel with the weights in equation (\ref{norm}).
The Moller operator converts the sum over $f$-states to
a sum over $u$-states, so we can simply add together the 
values at the energy eigenvalues shown in the figure.  
When this is done for $E_u<37,500$ 
cm$^{-1}$, the results for the four different scale
parameters give excellent agreement, particularly for $a=10-30$.
In the figure, all four curves clearly show the resonant
contribution as the lowest energy quadrature point.
The resolution of this resonant contribution is best
for the small scale parameters due to their closer
spacing of energy eigenvalues. The larger scale
parameters provide larger spacing which enables 
the quadrature to include higher energies.
In the present work, we found it convenient
to use $a=20$ for all IOS calculations.

Figure 5 shows partial TBR rate constants for
He+H$\cdot\cdot\cdot$H$(j=15)$ computed using the IOS approximation 
with the MR PES \cite{mr}.  Each curve includes a sum over all bound 
levels. The BAB resonance gives a contribution which is
comparable to the discretized states of the non-resonant continuum.
The low energy non-resonant states corresponding to $v=14-20$ give
contributions which range over several orders of magnitude, whereas
the higher energy $v=22-25$ states give contributions with less 
variation but with a gradual shift in threshold energy.
The sharp QB resonance gives more than an order of magnitude 
larger contribution than any of the other $j=15$ states.
This is due to the strong localization of the QB state at 
short distances and the neglible overlap with neighboring
continuum states (see Figure 3). 

In order to study the
effect of pressure on this QB contribution to the TBR
rate constant, we estimated the non-LTE concentration defect
$\delta_u$ given in equation (\ref{deltau}) using the IOS rate 
coefficients $k_{u\rightarrow b}$ and $k_{u\rightarrow u'}$ 
at $T$=1000 K. The maximum possible pressure-dependence
would occur when $\delta_b=-1$ which corresponds to zero
concentration of the bound level $b$. We assume $\delta_{u'}=0$ 
which is a good approximation for BAB resonances
and the non-resonant background \cite{pack2}. 
%For QB states where $\delta_{u'}\neq 0$, the rate
%coefficient connecting $u'$ with $u$ is typically
%small enough that the $\delta_{u'}=0$ assumption 
%would not substantially effect our estimate.
For QB states, the $\delta_{u'}=0$ assumption may be
used as the starting point for an iterative solution.
The result for $u=(13,15)$ is $\delta_u=-0.44$ in the 
[He]$\rightarrow\infty$ limit. This defect provides the
maximum amount of falloff with pressure that can occur 
for this QB state at the given temperature. Tables I and II 
show the maximum defects $\delta_u^{max}$ for all resonances
with tunneling widths less than 0.01 cm$^{-1}$.
Also given in the tables is the critical density of
He atom colliders defined by equation (\ref{cr1})
%\begin{equation}
%\mbox{[He]}_{cr}=\frac{1}{\tau_u\left(\sum_b k_{u\rightarrow b}
%+\sum_{u'\neq u} k_{u\rightarrow u'}\right)}
%\label{cr1}
%\end{equation}
which gives an estimate of where the falloff 
would be expected to occur. Not surprisingly,
the defects are largest in magnitude for the extremely 
long-lived states $u=(6,24),(6,29)$, and $(3,32)$
where the critical density is effectively zero.
However, even in these cases, the cancellation
in equation (\ref{krnew}) is not complete, and
the QB state contributes to the effective TBR 
rate constant. For the other QB states, the
typical value $\delta_u^{max}=-0.5$ shows that
the QB contribution would be at most reduced 
in half at high densities. 

It is also noteworthy that the high density limit tends to move 
the system toward the LTE limit, so the $\delta_b=-1$ assumption 
is generally too severe. The only way the bound states can remain
unpopulated is for there to exist an efficient mechanism, e.g. the
radiative contribution in equation (\ref{deltab}), which prevents 
three-body collisions from populating excited bound and QB states. 
If such a mechanism does not exist or is inefficient compared to 
the three-body collisions, then the $\delta_u^{max}$ may be 
substituted back into equations (\ref{deltau}) and (\ref{deltab}) 
and the system approaches the LTE limit after a few iterations.
Similar to the critical density (\ref{cr1}) for pressure falloff, 
we may use equation (\ref{cr2}) to determine a critical density 
%\begin{equation}
%\mbox{[C]}_{cr}=\frac{\sum_{b'} \Gamma_{b\rightarrow b'}}
%{\sum_u k_{b\rightarrow u}+\sum_{b'\neq b} k_{b\rightarrow b'}}
%\label{cr2}
%\end{equation}
for departures from LTE.
For homonuclear systems such as H$_2$, the inefficiency
of quadrupole radiation suggests that this critical
density should be small.
In fact, several master equation studies of CID for H$_2$ 
in astrophysical environments \cite{roberge,lepp,mandy} 
have shown that it is only at very low densities that non-LTE 
behavior would be important and the $\delta_b=-1$ assumption
would be valid, and in this case only for the highly excited states. 
The bound levels thermalize progressively 
with increasing gas density with the higher excited states 
thermalizing later due to the larger radiative transition 
probabilities. 
%Models of primordial star formation \cite{palla,flower}
Primordial star formation models \cite{palla,flower}
have shown that all bound and continuum states are thermalized
at densities around $10^{13}$ cm$^{-3}$. This density is less 
than many of the critical densities given in Tables I and II.
Therefore, the falloff due to the removal of QB state contributions 
from the pathway-independent part of the rate constant (\ref{krnew})
is generally small for an isolated H$_2$ system.

The results of the IOS approximation were benchmarked against the more 
accurate CS approximation to determine the temperature regime where 
they may be considered to be reliable. In both cases, the renormalized
Numerov method \cite{numerov} was used for propagation over $R$
with a step size of 0.01 a.u. and a matching radius of 50 a.u.
In the CS calculations, the PES was expanded in Legendre polynomials,
$P_{\lambda}(\cos\theta)$,
with truncation limit $\lambda_{max}=10$. Likewise, in the IOS
calculations, a Legendre expansion was used for the T-matrix 
with the same value of $\lambda_{max}$. A 40-point Gaussian quadrature 
was used to integrate over $\theta$ with only 20 angles needed
in the computations due to homonuclear symmetry. We used 
$l_{max}=10$ for collision energies between 1 and 10 cm$^{-1}$, 
$l_{max}=30$ for energies between 10 and 100 cm$^{-1}$,
$l_{max}=60$ for energies between 100 and 1,000 cm$^{-1}$, 
$l_{max}=120$ for energies between 1,000 and 10,000 cm$^{-1}$, and
$l_{max}=200$ for energies between 10,000 and 100,000 cm$^{-1}$.
This wide energy grid and a small energy step-size ensured that 
the Boltzman average in equation (\ref{kij}) contained neglible 
interpolation and trunction error.

For the CS calculations performed in the present work,
we extended the temperature range of previous calculations
\cite{paolini} using the scale parameters given by
Ohlinger et al. \cite{luke}. 
These scale factors increase with $j$                
in order to get a good representation of the QB states
and to increase the spacing between the positive energies.
This reduces the amount of vibrational coupling and allows
the calculations to be more tractible, however, it introduces
some numerical error in the Sturmian evaluation of $Q_T$ (see Figure 1a)
which limits the reliability of the results, particularly 
at low temperatures. 
In the IOS calculations, the vibrational motion was decoupled
from the dynamics, so the efficiency of the computations did not
depend on the choice of scale parameter. Therefore, we were able to
choose a smaller value and effectively remove this source of error from the 
calculations.
Figure 6 shows that the CS and IOS results agree very well for
temperatures greater than 600 K. In both sets of calculations,
the basis sets were restricted to $j_{max}=20$. For larger
$j$-values, the CS calculations become inefficient
due to the increased coupling and larger number of 
projection quantum numbers. The IOS results for $j_{max}=30$
illustrate the importance of the larger $j$-values as the temperature
is increased. Although theses results are very nearly converged over
the entire temperature range shown, the apparent agreement with the 
experimental data point \cite{trainor} at 300 K is not meaningful. 
The CS curve does appear to give a similar temperature dependence 
as the experimental data but with a larger magnitude. Increasing
the value of $j_{max}$ for the CS calculations would further 
increase this discrepancy.

The above analysis shows that pressure falloff cannot be the
source of the disagreement.  As was noted previously \cite{paolini}, 
there is significant uncertainty in the MR PES when the H-H bond is 
stretched. This uncertainty can have an affect on the TBR rate constant
and is the most likely source of the discrepancy between the CS results 
and the experiment.  In order to further explore this possibility, we tested
a pairwise additive PES consisting of the He-H potential \cite{nicolais}
\begin{equation}
v(r)=2.2\times 10^{-5}e^{-0.8(r-6.8)}\left[e^{-0.8(r-6.8)}-2\right]
\ \ \ \mbox{a.u.}
\end{equation}
using the IOS approximation.  For an inert collider such as He, 
it has been argued that a pairwise additive PES would give a reasonable 
estimate of the TBR rate \cite{rbc}. The result is shown in Figure 7 
which compares the TBR rate constants computed with the two surfaces. 
The results do not agree particularly well at temperatures less than 10,000 K
which confirms that the uncertainty in the PES is largely responsible
for the discrepancy with experiment. 
Unfortunately, it is not clear to what extent the sensitivity is due to
the missing three-body terms in the pairwise additive PES or the uncertainty
in the MR PES when the H-H bond is stretched.  It would be desirable to have an 
improved PES for this system before undertaking any further computationally 
expensive CS or CC calculations. The uncertainties in the rate constants
for He colliders, however, do not play a major role in limiting the reliability of
simulations of H$_2$ formation in primordial gas due to the relatively 
low density of He. The major source of uncertainty is due to H colliders.

For H+H+H, the same basis set and numerical parameters were used
as for He colliders. Therefore, the IOS approximation is estimated to be 
reliable above 600 K (see Figure 6). The distinguishable particle cross sections
were multiplied by three in order to account for the three possible pairs of
molecules that may recombine.  The results are shown in Figure 8 along with 
a fit to the experimental data of Jacobs et al. \cite{jacobs} and an extrapolation 
which is frequently used in astrophysical models \cite{turk}. The present
results are in reasonable agreement at high temperature and are
within a factor of 2 of the experiment \cite{jacobs}.
However, we find a much flatter temperature dependence with decreasing
temperature than is given by the extrapolation.
The present results are also similar to the DEB quasiclassical calculations 
of Esposito and Capitelli \cite{esposito}. 
The similarity between the classical result
and the quantum mechanical result is very
encouraging and shows that the temperature dependent
fit \cite{jacobs} does not reliably extrapolate to
lower temperatures. This conclusion is further
strengthened by the insensitivity of the classical
calculations \cite{esposito} to the BKMP2 versus LSTH
surfaces. Therefore, the uncertainty in the PES which
plagued the TBR calculations for He colliders is not
applicable for H colliders.  It is also noteworthy that 
the experimental data \cite{jacobs} were based on shock tube measurements 
which  were not claimed to be of high accuracy. In fact, the results
were in the middle of a range of reported values that
scattered over an order of magnitude. Considering this
uncertainty and the approximate treatment of the dynamics
in the theoretical calculations, it is difficult to say
which result is the most reliable. Nevertheless, 
some of the uncertainty in the TBR rate constant used 
in astrophysical models may be removed. The application 
of detailed balance to the phenomenological rate constants 
$k_r$ and $k_d$ is clearly defined in the Sturmian theory
which allows statistical errors to be easily identified 
and removed from consideration \cite{bob}.
Large uncertainties associated with the various
extrapolation methods \cite{turk,bob}
may be replaced by smaller uncertainties associated
with the dynamical approximations.
The present results, the experimental data \cite{jacobs},
and the classical DEB results \cite{esposito}, are all within 
a factor of 2 for temperatures in the experimental range 2900-4700 K, 
and the two theoretical calculations are within a factor of 2 for all 
temperatures above 300 K. Therefore, the factor of $\sim$ 100
uncertainty which was introduced by previous rate constants 
\cite{turk} is estimated to be reduced to a factor of $\sim$ 2 
when either the present or DEB results are used in the temperature 
range required by the astrophysical simulations \cite{turk,bovino}.

\section{Conclusions}

In the well-known resonance theory of molecule formation, 
the main quantum feature is the identification of the 
appropriate collision complex \cite{rbc}.
The QB orbiting resonances are generally used to identify
two-step mechanisms for recombination, and classical or 
quantum mechanical calculations are then used to describe
the dynamics. Results from the different two-step mechanisms
are then typically \cite{whitlock,orel} added together 
to obtain the total TBR rate.
Adding the various contributions together may lead to double counting
of the kind pointed out by Wei, Alavi, and Snider \cite{wei} who showed
that each mechanism must give the same result for systems at equilibrium
when the calculations are carried out exactly.

Although the ORT is still in wide use, it has been shown that
non-resonant processes are generally not negligible, and in many instances 
provide the dominant contribution to the TBR rate \cite{pack2,schwenke}.
The quantum kinetic theory of Snider and Lowry \cite{snider} generalizes
ORT to include non-resonant states, but difficulties in its implementation \cite{pack1}
have prevented it from being used in practice. The Sturmian theory provides 
a practical implementation of the quantum kinetic theory \cite{snider}
and generalizes it further to include metastable states that are
formed as independent species.  Unlike ORT which uses energy and 
lifetime considerations to select the most important QB states 
(sometimes called RBC states) for recombination, the Sturmian theory 
retains all unbound states for a given pathway.
Calculations are performed for a single pathway only, and
the result is considered to be the complete TBR rate in the
LTE limit. For non-LTE systems, the QB states may be treated
as distinct species and the pathway dependence of their
formation may be incorporated via a set of rate equations 
which maintain the basic structure of the quantum kinetic 
theory \cite{snider}. Closed-form expressions for the non-LTE
solutions are easily derived in terms of lifetimes, rate coefficients,
and number density.  The non-LTE corrections are shown to be
small for systems which do not have an efficient mechanism 
for depopulating the excited bound and QB states.
In the absence of such a mechanism, 
the pathway-independent part is sufficient to 
calculate the rate constants, and the result
does not depend on pressure as is commonly
assumed in ORT. 

The Sturmian theory eliminates uncertainties associated with
pathway dependence and ensures that all contributions are accounted 
for exactly once. The quantum calculations of Pack, Walker, and 
Kendrick \cite{pack1} do not suffer the double counting problem,
however, their interpretation of the various mechanisms differs 
from \cite{wei} in that it does not distinguish between the free 
continuum and the BAB resonant and non-resonant part of the
interacting continuum.  
This precludes the use of the quantum kinetic theory \cite{snider}
and requires a numerical solution to the master equations with 
a subsequent fit to the effective rate equation (\ref{rate}).
Nevertheless, a model based on the results of their master equation 
analysis shows very similar behavior to the Sturmian theory given here. 
The classical calculations of Esposito and Capitelli \cite{esposito} 
also do not distinguish between the free continuum and the BAB 
resonant and non-resonant part of the interacting continuum. 
These calculations added together a recombination piece obtained 
from CID using detailed balance, their so-called DEB result, 
with a separate calculation of the ORT contribution.
We interpret their DEB result as a classical calculation
of the pathway-independent part of the rate constant
(computed using the direct mechanism, $k_r^0$) and argue
that it should not be added to the ORT contribution.
The ORT contribution is relatively small in the temperature 
range of the experiments \cite{jacobs}, and the factor of two 
agreement between the present quantum results, the classical results, 
and the experimental data is encouraging.

The ET mechanism was used in the present work, however,
alternate implementations are certainly possible.  
For example, if one wished to formulate the theory
in terms of the direct process (\ref{1a}), then the
Sturmian representation used in the dynamics would need 
to be for the free $AB$ continuum rather than the interacting
$A\cdot\cdot\cdot B$ subsystem used here. The dynamical
calculations would perhaps be more difficult to solve
since the transition operator $T_0$ acting on the
free Sturmian basis set would not be of the usual $V+VGV$ form.
Likewise, the exchange process (\ref{1c}) would require a 
Sturmian representation of the $A\cdot\cdot\cdot C$ 
subsystem followed by a dynamical calculation of the
rearrangement, presumably a calculation of greater 
difficulty than the single arrangement dynamics 
considered here.  

The present numerical study has shown how to apply the Sturmian theory
at relatively high temperatures where multiple scattering and interference
effects are largely neglible. These calculations are sufficient to address
uncertainties that have limited the reliability of astrophysical simulations 
of primordial star formation \cite{turk}. The same framework,
however, should be applicable at intermediate temperatures with an improved
quantum mechanical description of the dynamics.
The simplified IOS approximation \cite{pfeffer} may be extended to include 
vibrational coupling in the dynamical calculation \cite{parker}. 
For the H+H+H system, it is likely that all of the 101 vibrational functions 
used in the Sturmian represention would need to be coupled together 
for each orientation angle.  At low temperatures, the methods described 
here would need to be further developed. The IOS approximation breaks down     
and should be replaced by the CS approximation or even the 
full CC formulation. For systems which contain an inert gas, 
such as He+H+H and Ar+H+H, the CS approximation has been used 
with a Sturmian representation of the ET mechanism to describe
TBR \cite{paolini}. The results appear to give the correct
temperature dependence but are inconclusive due 
to uncertainties in the potential energy surfaces. For
systems which support bound states in more than one
arrangement, there would likely be interference
effects at low temperatures,
%between the different mechanisms at low temperatures, 
and a more sophisticated quantum dynamical calculation 
\cite{parker2,parker3,esry1,esry2,esry3}
would be required.

\begin{center}
{\bf Acknowledgments}
\end{center}

This work was supported by the NSF grant No. PHY-1203228. The
author would like to thank Dr. Brian Kendrick for a careful
reading of the manuscript and Dr. Nicolais Guevara for providing
the He-H potential used in this work.

\newpage
\centerline{\psfig{figure=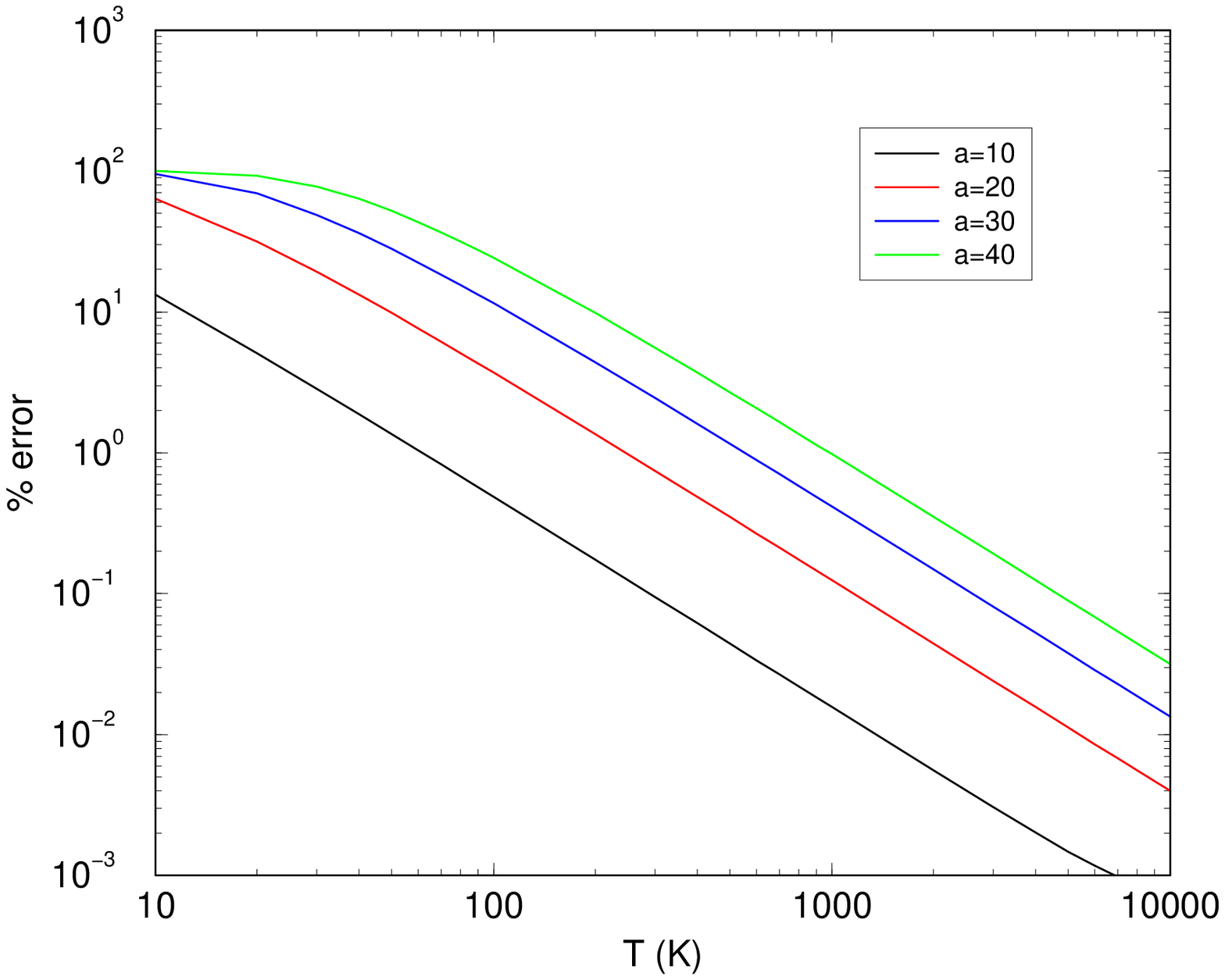,width=3in}\hspace{.1in}
\psfig{figure=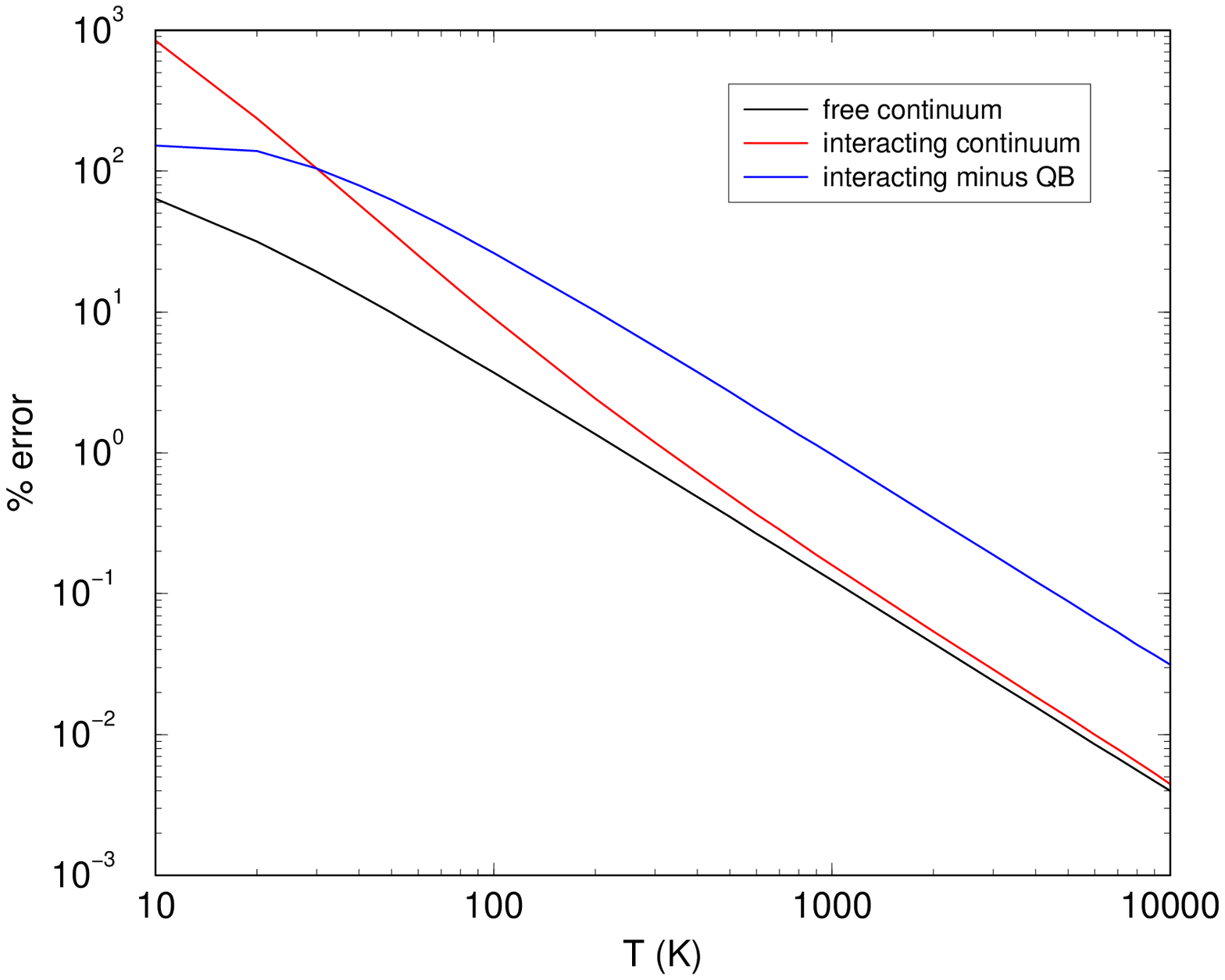,width=3in}}
\vspace*{.2in}
\footnotesize{\noindent Figure 1: (Color online) Percent error in the translational 
partition function $Q_T$ computed using the Sturmian eigenvalues for $j=4$.
(a) Free particle interaction with different values
of scale parameter; (b) Interacting continuum with and without
the long-lived QB state included in the representation. 
These calculations used $a=20$.}

\newpage
\centerline{\psfig{figure=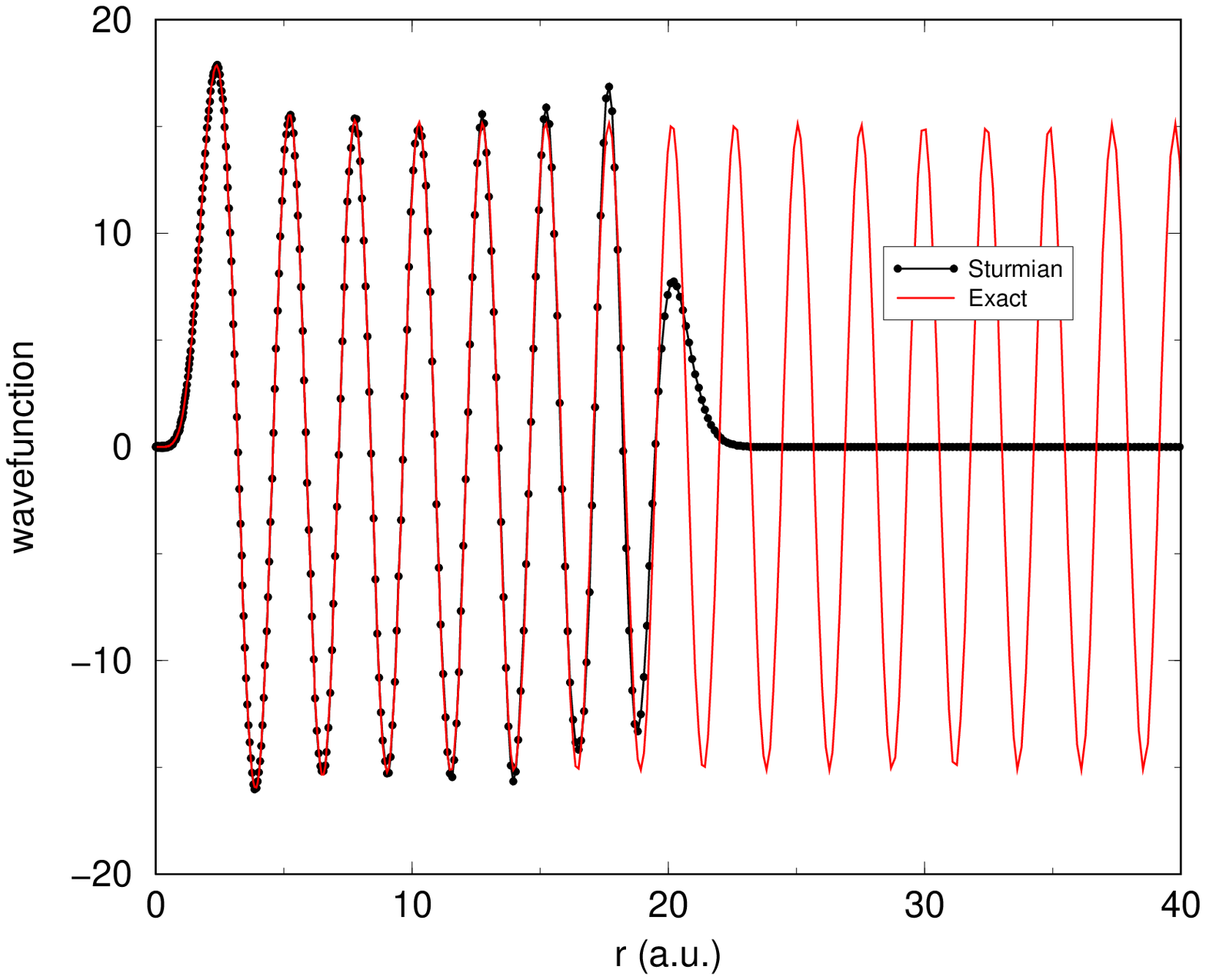,width=3in}\hspace{.1in}
\psfig{figure=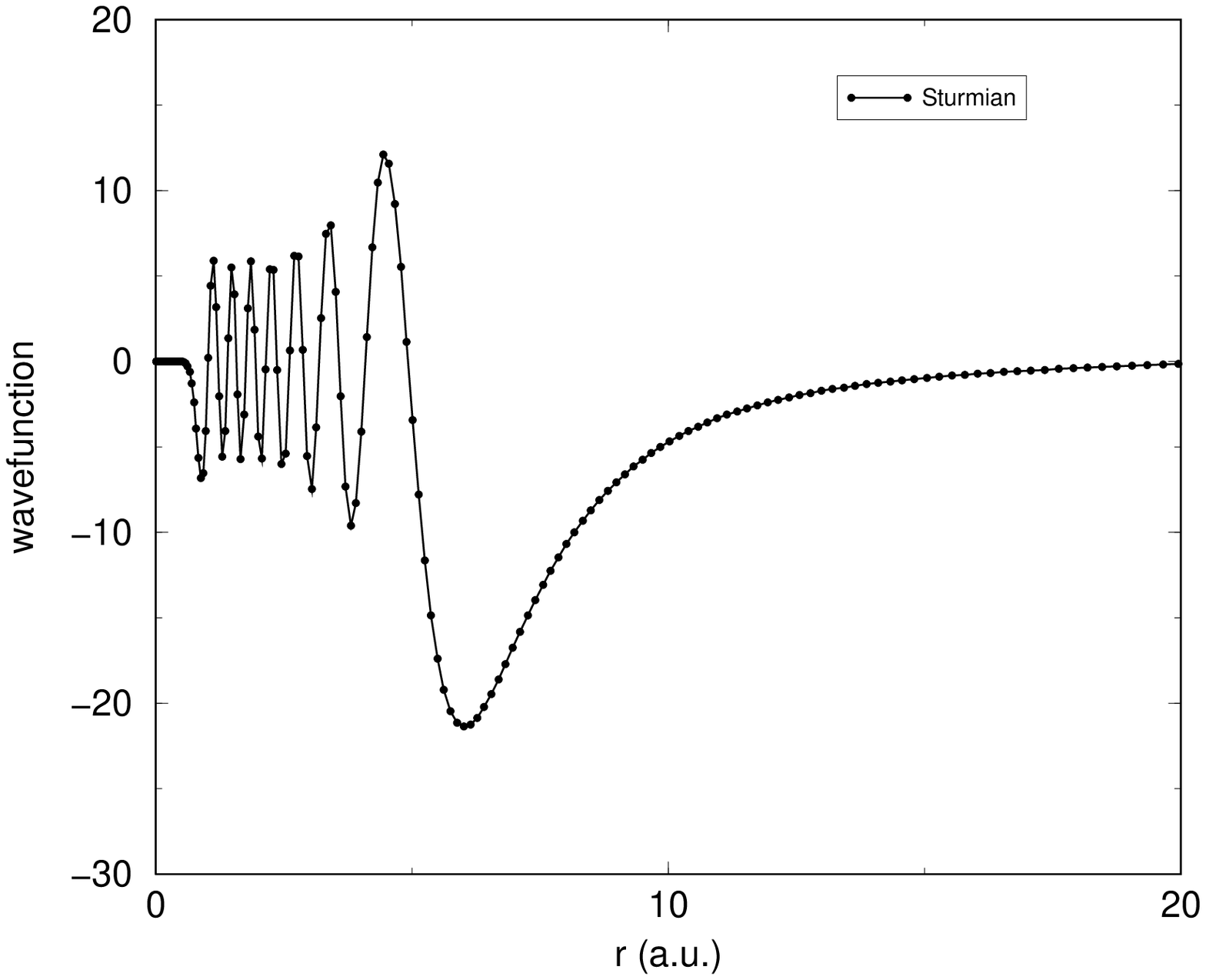,width=3in}}
\vspace*{.2in}
\footnotesize{\noindent Figure 2: (Color online) Sturmian representation of 
$j=4$ eigenstates for (a) free continuum and (b) interacting 
continuum. In both cases, the 14th vibrational eigenstate 
is plotted for $a=20$. This choice of scale factor causes
all of the continuum eigenstates to go to zero for interatomic
distances greater than 25 a.u.  The interacting eigenstate
corresponds to a QB resonance with an energy of $0.9$ cm$^{-1}$ 
and a width of $8.4\times 10^{-6}$ cm$^{-1}$.
}

\newpage
\centerline{\psfig{figure=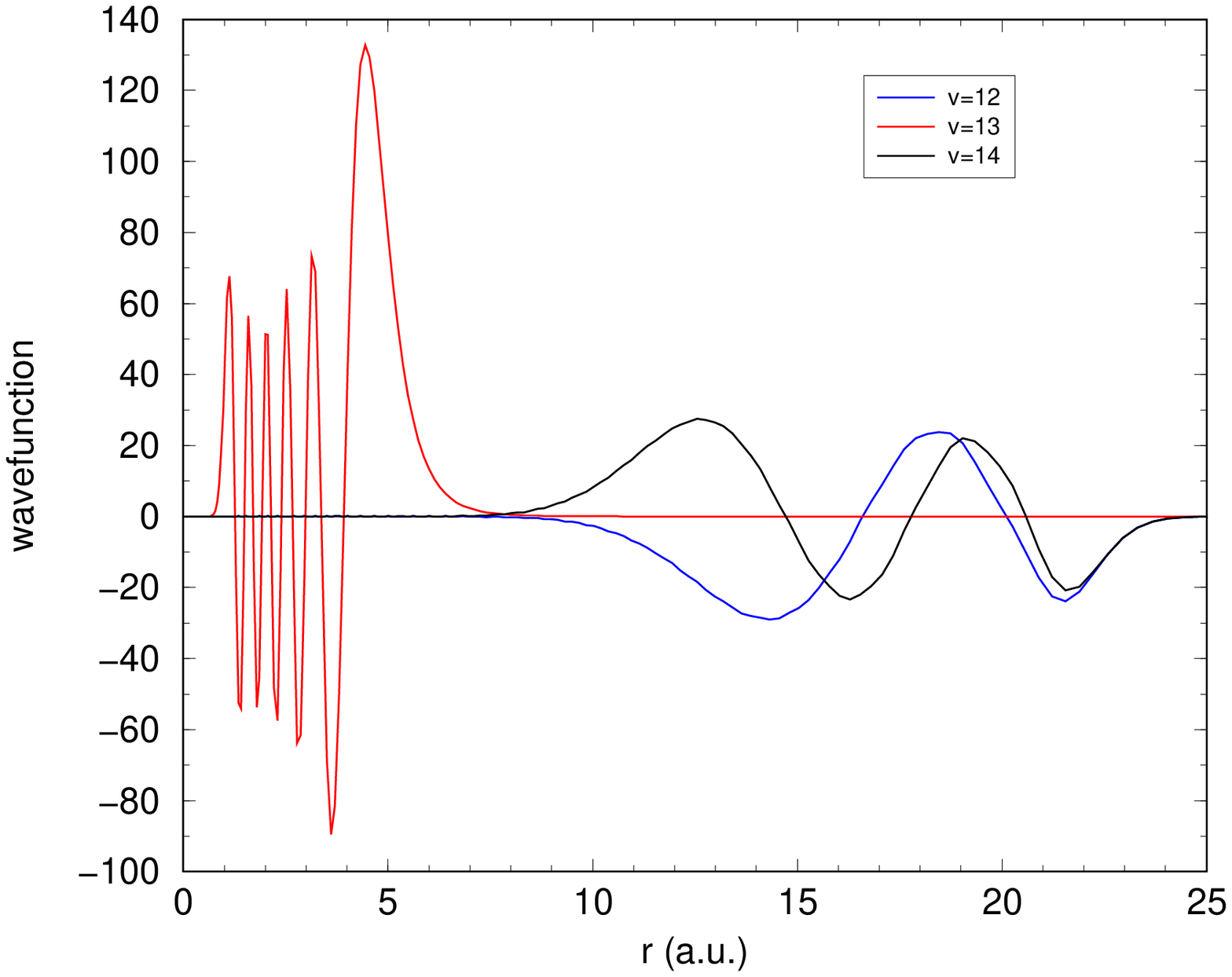,width=6in}}
\vspace*{.2in}
\footnotesize{\noindent Figure 3: (Color online) Sturmian representation of 
the $j=15$ interacting continuum eigenstates for $v=12-14$ using
a scale factor $a=20$.
The $v=13$ eigenstate corresponds to a QB resonance with
a width of $3.2\times 10^{-6}$ cm$^{-1}$. The figure shows
there is neglible coupling between the resonance
and the neighboring continuum states. Higher vibrational levels
do show overlap and can help keep the QB concentration from 
being depleted.
}

\newpage
\centerline{\psfig{figure=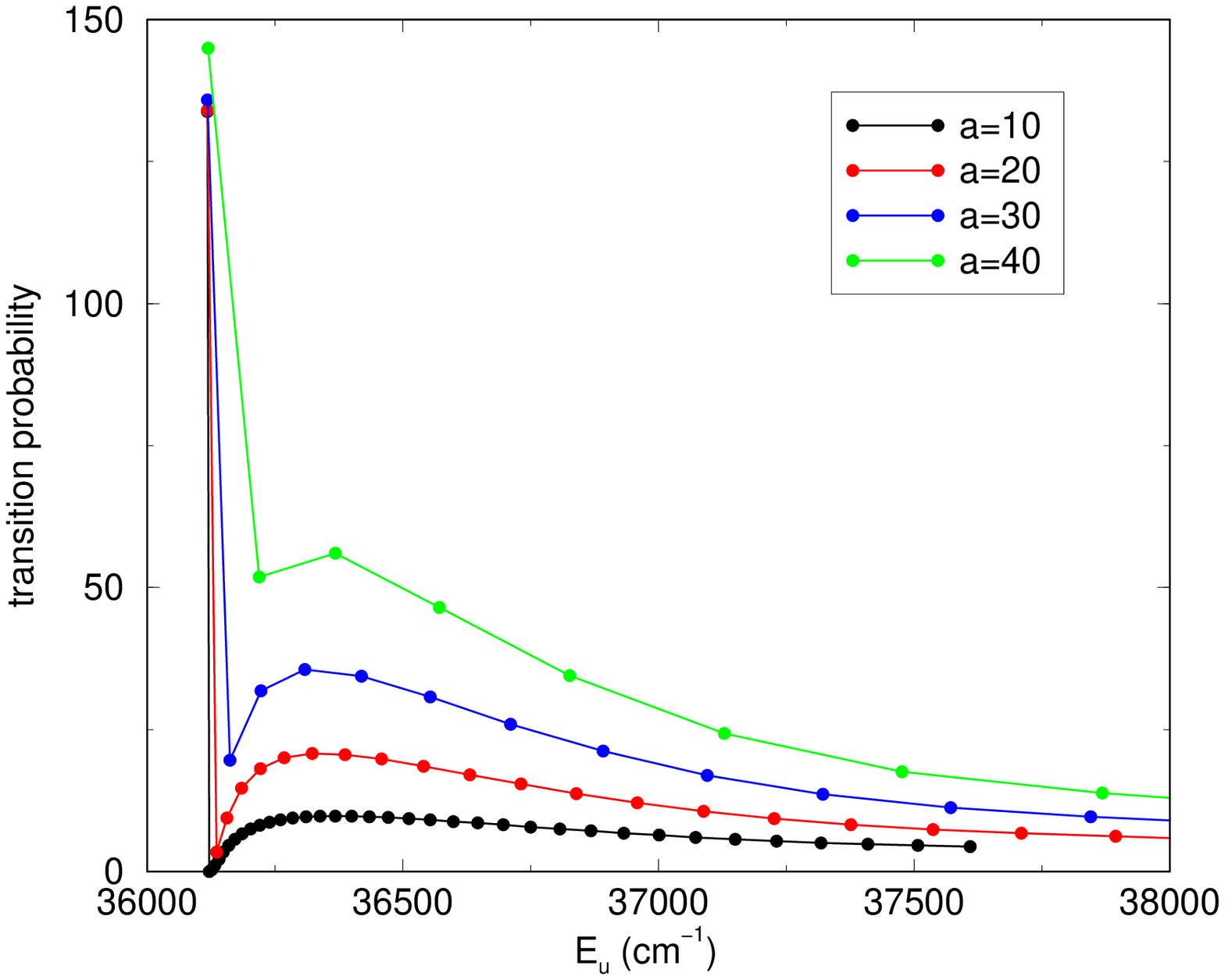,width=6in}}
\vspace*{.2in}  
\footnotesize{\noindent Figure 4: (Color online) Cumulative bound-free transition 
probability as a function of Sturmian energy and scale parameter. The calculations
are for He+H$\cdot\cdot\cdot$H$(j=4)$ computed with the MR PES. 
The scale dependence is due to the unit normalization of the Sturmian
basis set. The energy normalization cancels with the equivalent
quadrature weights for the sum over $f$ and allows the sum over
$u$ to be obtained by simply adding the values at the points 
shown in the figure. For example, the sum over $u$ for 
$E_u<37,500$ cm$^{-1}$ yields 366.49,  366.43,  365.93, 
and 375.72 for the respective curves $a$=10, 20, 30, and 40.
The IOS approximations was used for these calculations with 
$E_T=100$ cm$^{-1}$ and $\overline{l}_{max}=30$.
}

\newpage
\centerline{\psfig{figure=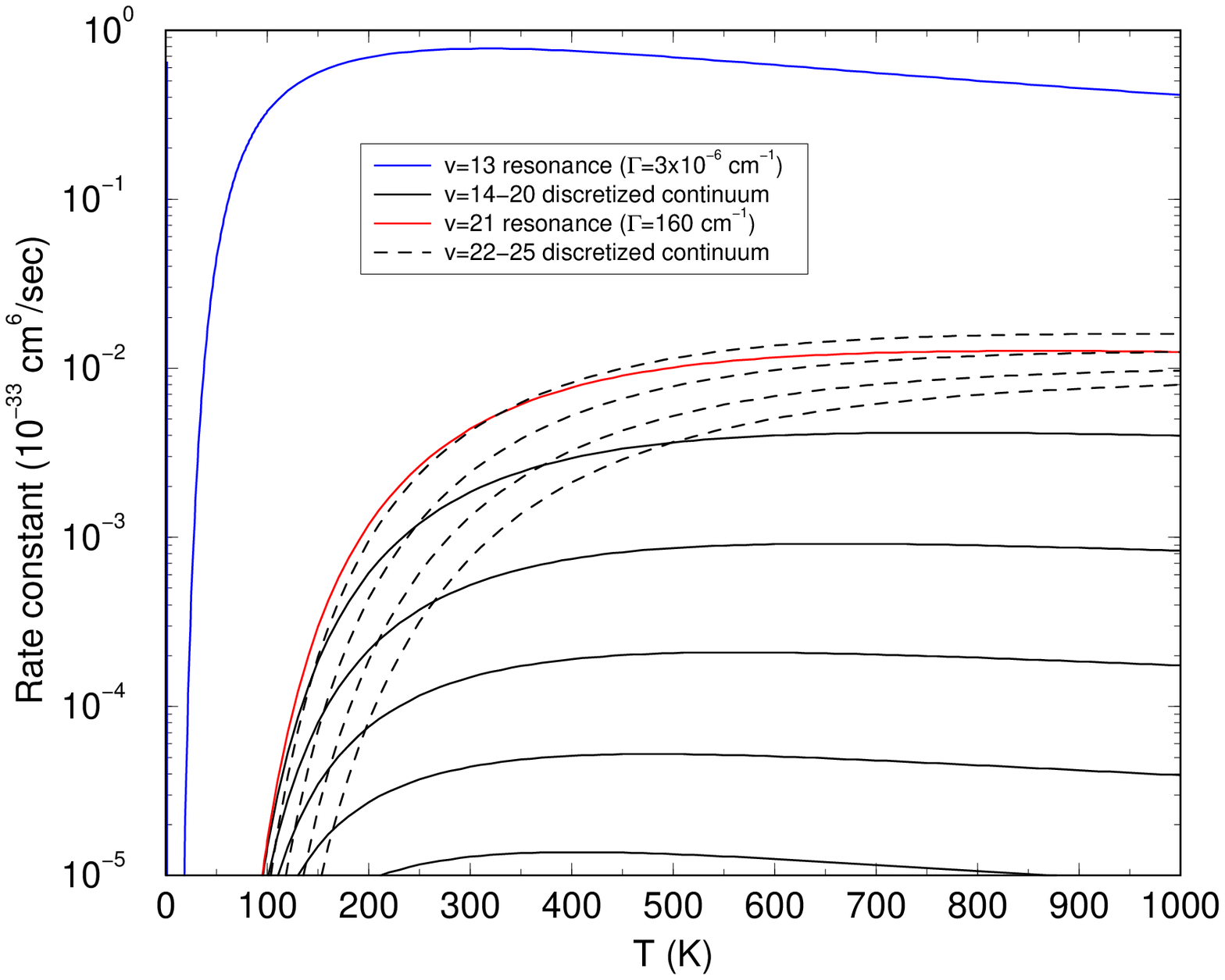,width=6in}}
\vspace*{.2in}
\footnotesize{\noindent Figure 5: (Color online) TBR partial rate constants 
for He+H$\cdot\cdot\cdot$H$(j=15)$ computed with the MR PES. Each
curve includes a sum over all bound levels. The resonant 
contributions are shown in blue (QB) and red (BAB), 
and the non-resonant contributions are shown as solid lines 
(low energies) and dashed lines (high energies).
}

\newpage
\centerline{\psfig{figure=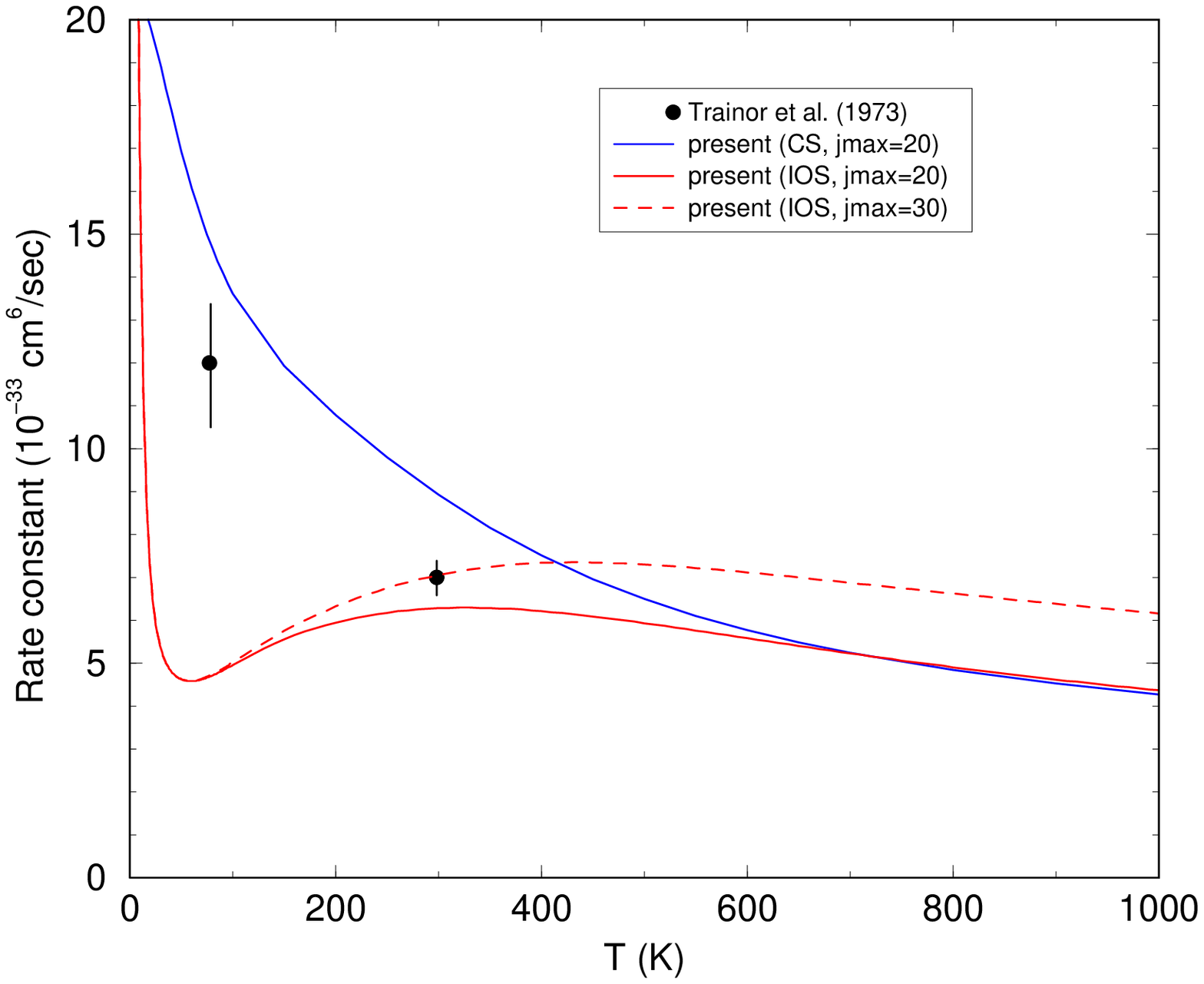,width=6in}}
\vspace*{.2in} 
\footnotesize{\noindent Figure 6: (Color online) TBR rate constant for He+H+H 
computed with the MR PES. The CS result appears to give a similar 
temperature dependence as the experimental data of Trainor et al. \cite{trainor}
but with a larger magnitude. When the same $j_{max}=20$ 
condition is used, the IOS result and the CS result show good agreement 
for temperatures above 600 K.  The IOS result for $j_{max}=30$ shows that
additional $j$-values are needed to achieve convergence for $T>100$ K.
}

\newpage
\centerline{\psfig{figure=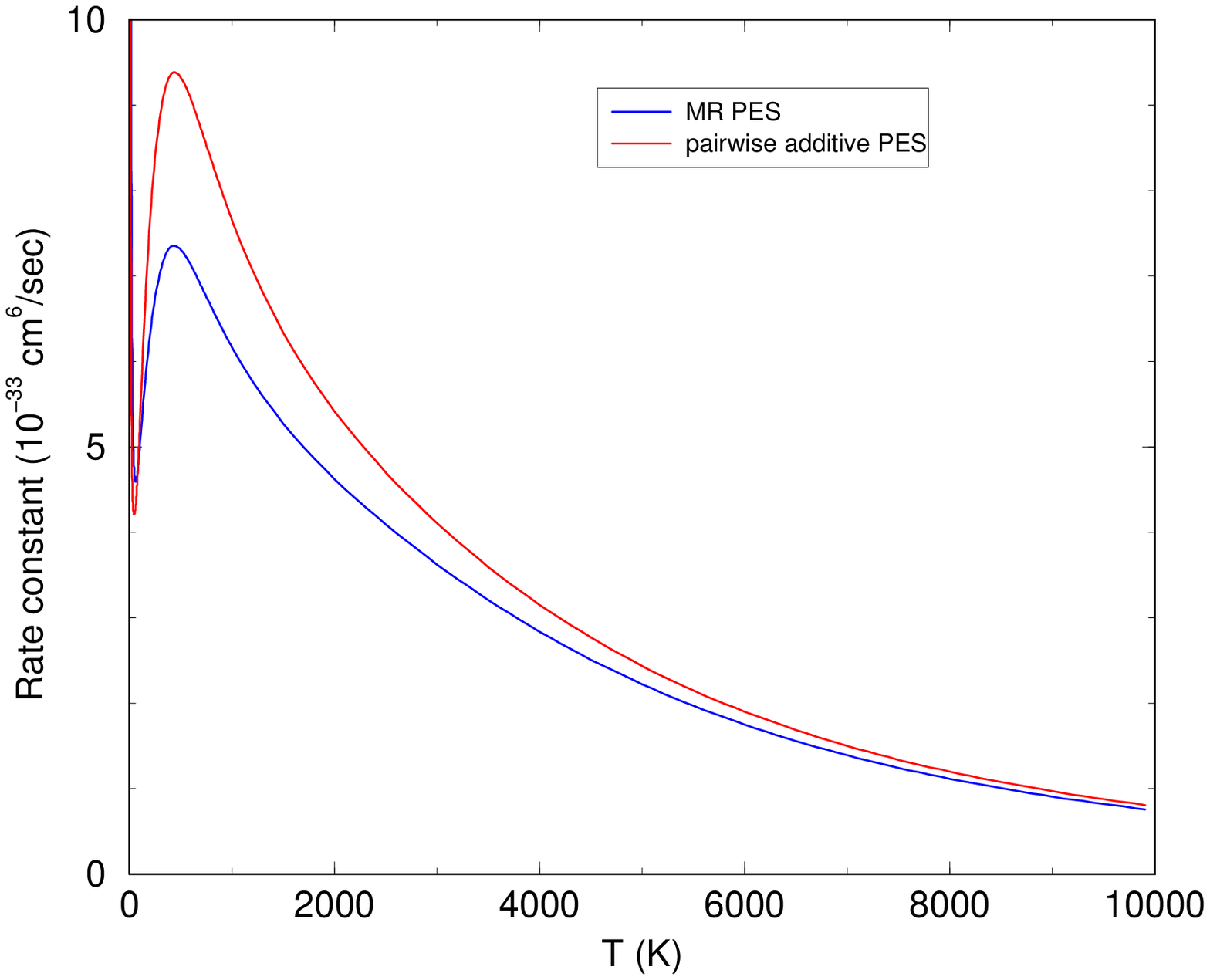,width=6in}}
\vspace*{.2in}
\footnotesize{\noindent Figure 7: (Color online) TBR rate constant for He+H+H
computed with the MR PES \cite{mr} and pairwise additive PES \cite{nicolais}.
Both calculations used the IOS approximation with $j_{max}=30$.
The uncertainty in the MR PES at large-$r$ and the missing three-body 
terms in the pairwise additive PES are responsible for the significant 
difference in the TBR rate for $T<10,000$ K.
}

\newpage
\centerline{\psfig{figure=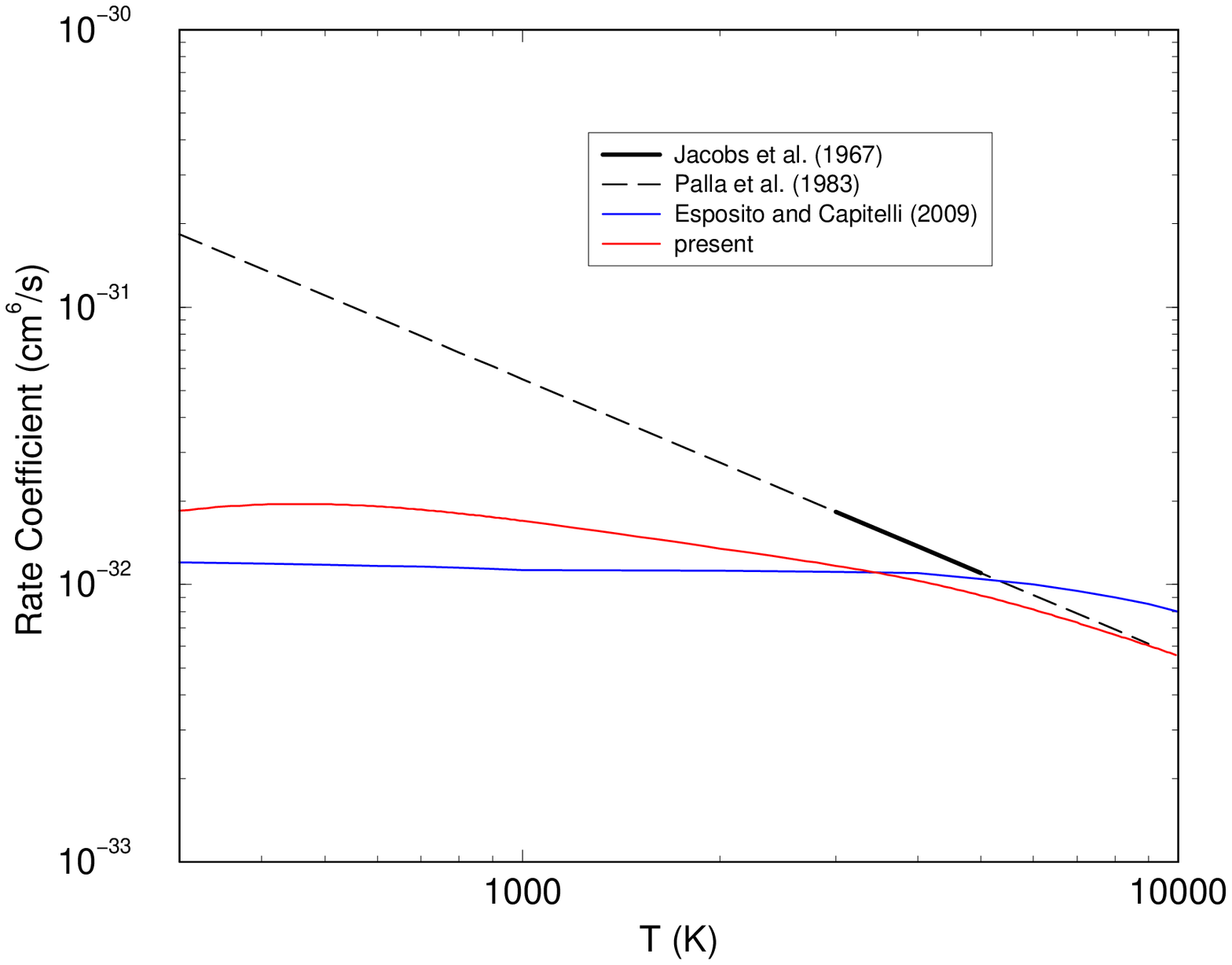,width=6in}}
\vspace*{.2in}
\footnotesize{\noindent Figure 8: (Color online) TBR rate constant for H+H+H
computed with the BKMP2 PES using the IOS approximation with 
$j_{max}=35$. The present results and the quasiclassical DEB
results of Esposito and Capitelli \cite{esposito} show a much 
flatter temperature dependence than the extrapolation (dashed curve) 
which is based on a fit to the experimental data of Jacobs et al \cite{jacobs}
in the temperature range 2900-4700 K. 
}

\newpage
\begin{table}[h]
\centering
\caption{Resonance parameters for para-H$_2 (X^1\Sigma_g^+)$ at $T=1000$ K.
The QB lifetimes were obtained from the widths reported by 
%Schwenke \cite{schwenke2},
Schwenke [33]
and the rate coefficients for collision with He were computed using 
%the MR PES \cite{mr}.
the MR PES [34].
}
\vspace*{.2in}
\begin{tabular}{|c|c|c|c|c|c|}\hline
\multicolumn{1}{|c|}{\hspace*{.1in}$u=(v,j)$\hspace*{.1in}} &
\multicolumn{1}{c|}{$\tau_u$ (s)\hspace*{.1in} } &
\multicolumn{1}{c|}{$\sum_b k_{u\rightarrow b}$ (cm$^3$s$^{-1}$)\hspace*{.1in}} &
\multicolumn{1}{c|}{$\sum_{u'} k_{u\rightarrow u'}$ (cm$^3$s$^{-1}$)\hspace*{.1in}} &
\multicolumn{1}{c|}{[He]$_{cr}$ (cm$^{-3}$)\hspace*{.1in} } &
\multicolumn{1}{c|}{$\delta_u^{max}$\hspace*{.1in}}\\ \hline
(14,4) & 6.3$\times 10^{-7}$\hspace*{.1in} & 1.3$\times 10^{-9}$ & 4.0$\times 10^{-9}$ &
3.0$\times 10^{14}$ & -0.25 \hspace*{.1in} \\ \cline{1-6}
(6,24) & 4.8$\times 10^{9}$\hspace*{.1in} & 1.0$\times 10^{-8}$ & 4.7$\times 10^{-9}$ &
0 & -0.69 \hspace*{.1in} \\ \cline{1-6}
(10,26) & 2.8$\times 10^{3}$\hspace*{.1in} & 8.2$\times 10^{-9}$ & 7.0$\times 10^{-9}$ &
2.4$\times 10^{4}$ & -0.54 \hspace*{.1in} \\ \cline{1-6}
(14,28) & 2.5$\times 10^{0}$\hspace*{.1in} & 8.0$\times 10^{-9}$ & 7.3$\times 10^{-9}$ &
2.6$\times 10^{7}$ & -0.52 \hspace*{.1in} \\ \cline{1-6}
(17,30) & 3.8$\times 10^{-2}$\hspace*{.1in} & 8.0$\times 10^{-9}$ & 6.1$\times 10^{-9}$ &
1.9$\times 10^{9}$ & -0.57 \hspace*{.1in} \\ \cline{1-6}
(3,32) & 8.8$\times 10^{21}$\hspace*{.1in} & 8.7$\times 10^{-9}$ & 1.2$\times 10^{-9}$ &
0 & -0.88 \hspace*{.1in} \\ \cline{1-6}
(21,32) & 3.0$\times 10^{-3}$\hspace*{.1in} & 7.7$\times 10^{-9}$ & 5.5$\times 10^{-9}$ &
2.6$\times 10^{10}$ & -0.58 \hspace*{.1in} \\ \cline{1-6}
(24,34) & 9.7$\times 10^{-4}$\hspace*{.1in} & 2.5$\times 10^{-9}$ & 9.3$\times 10^{-9}$ &
8.8$\times 10^{10}$ & -0.21 \hspace*{.1in} \\ \cline{1-6}
\end{tabular}
\end{table}

\newpage
\begin{table}[h]
\centering
%\caption{Resonance parameters for ortho-H$_2 (X^1\Sigma_g^+)$ at $T=1000$ K.}
\caption{Same caption as Table I but for ortho-H$_2 (X^1\Sigma_g^+)$ at $T=1000$ K.}
\vspace*{.2in}
\begin{tabular}{|c|c|c|c|c|c|}\hline
\multicolumn{1}{|c|}{\hspace*{.1in}$u=(v,j)$\hspace*{.1in}} &
\multicolumn{1}{c|}{$\tau_u$ (s)\hspace*{.1in} } &
\multicolumn{1}{c|}{$\sum_b k_{u\rightarrow b}$ (cm$^3$s$^{-1}$)\hspace*{.1in}} &
\multicolumn{1}{c|}{$\sum_{u'} k_{u\rightarrow u'}$ (cm$^3$s$^{-1}$)\hspace*{.1in}} &
\multicolumn{1}{c|}{[He]$_{cr}$ (cm$^{-3}$)\hspace*{.1in} } &
\multicolumn{1}{c|}{$\delta_u^{max}$\hspace*{.1in}}\\ \hline
(14,13) & 1.3$\times 10^{-9}$\hspace*{.1in} & 5.2$\times 10^{-9}$ & 7.8$\times 10^{-9}$ &  
5.9$\times 10^{16}$ & -0.40 \hspace*{.1in} \\ \cline{1-6}
(13,15) & 1.7$\times 10^{-6}$\hspace*{.1in} & 6.1$\times 10^{-9}$ & 7.7$\times 10^{-9}$ &  
4.3$\times 10^{13}$ & -0.44 \hspace*{.1in} \\ \cline{1-6}
(12,17) & 1.5$\times 10^{-4}$\hspace*{.1in} & 6.6$\times 10^{-9}$ & 7.9$\times 10^{-9}$ &  
4.6$\times 10^{11}$ & -0.46 \hspace*{.1in} \\ \cline{1-6}
(12,19) & 4.1$\times 10^{-4}$\hspace*{.1in} & 7.0$\times 10^{-9}$ & 8.3$\times 10^{-9}$ &  
1.6$\times 10^{11}$ & -0.46 \hspace*{.1in} \\ \cline{1-6}
(13,21) & 1.4$\times 10^{-4}$\hspace*{.1in} & 7.3$\times 10^{-9}$ & 8.6$\times 10^{-9}$ &  
4.5$\times 10^{11}$ & -0.46 \hspace*{.1in} \\ \cline{1-6}
(15,23) & 2.4$\times 10^{-5}$\hspace*{.1in} & 7.6$\times 10^{-9}$ & 8.9$\times 10^{-9}$ &  
2.5$\times 10^{12}$ & -0.46 \hspace*{.1in} \\ \cline{1-6}
(17,25) & 4.1$\times 10^{-6}$\hspace*{.1in} & 7.8$\times 10^{-9}$ & 8.6$\times 10^{-9}$ &  
1.5$\times 10^{13}$ & -0.47 \hspace*{.1in} \\ \cline{1-6}
(20,27) & 7.7$\times 10^{-7}$\hspace*{.1in} & 7.8$\times 10^{-9}$ & 8.6$\times 10^{-9}$ &  
7.9$\times 10^{13}$ & -0.47 \hspace*{.1in} \\ \cline{1-6}
(6,29) & 6.5$\times 10^{12}$\hspace*{.1in} & 1.0$\times 10^{-8}$ & 3.0$\times 10^{-9}$ &  
0 & -0.78 \hspace*{.1in} \\ \cline{1-6}
(22,29) & 2.0$\times 10^{-7}$\hspace*{.1in} & 6.9$\times 10^{-9}$ & 9.5$\times 10^{-9}$ &  
3.0$\times 10^{14}$ & -0.42 \hspace*{.1in} \\ \cline{1-6}
(11,31) & 3.5$\times 10^{6}$\hspace*{.1in} & 8.2$\times 10^{-9}$ & 4.1$\times 10^{-9}$ &  
2.3$\times 10^{1}$ & -0.67 \hspace*{.1in} \\ \cline{1-6}
(25,31) & 8.0$\times 10^{-8}$\hspace*{.1in} & 3.7$\times 10^{-9}$ & 1.2$\times 10^{-8}$ &  
7.8$\times 10^{14}$ & -0.23 \hspace*{.1in} \\ \cline{1-6}
(16,33) & 6.0$\times 10^{3}$\hspace*{.1in} & 8.2$\times 10^{-9}$ & 2.6$\times 10^{-9}$ &  
1.6$\times 10^{4}$ & -0.76 \hspace*{.1in} \\ \cline{1-6}
(27,33) & 5.3$\times 10^{-8}$\hspace*{.1in} & 2.1$\times 10^{-9}$ & 1.2$\times 10^{-8}$ &  
1.3$\times 10^{15}$ & -0.15 \hspace*{.1in} \\ \cline{1-6}
(30,35) & 6.4$\times 10^{-8}$\hspace*{.1in} & 1.8$\times 10^{-9}$ & 1.1$\times 10^{-8}$ &  
1.2$\times 10^{15}$ & -0.14 \hspace*{.1in} \\ \cline{1-6}
\end{tabular}
\end{table}

\end{document}